\documentclass[twocolumn]{aastex62}
\usepackage{epsfig,amsmath,amsfonts,amssymb,graphicx,subfigure,enumitem,url,natbib}
\received{.........}
\revised{.........}
\accepted{........}
\submitjournal{AAS}
\shorttitle{Short period variable stars}
\shortauthors{Sneh Lata et al.}

\begin{document}
\title{Short period variable stars in young open cluster Stock 8}

\email{sneh@aries.res.in}

\author{Sneh Lata}
\affil{Aryabhatta Research Institute of Observational Sciences, Manora Peak, Nainital 263002, Uttarakhand, India}, 
\author{Anil. K. Pandey}
\affil{Aryabhatta Research Institute of Observational Sciences, Manora Peak, Nainital 263002, Uttarakhand, India}, 
\author{Ram Kesh Yadav}
\affil{National Astronomical Research Institute of Thailand, 191 Siriphanich Bldg., Huay Kaew Rd., Suthep, Muang, Chiang Mai
50200, Thailand}, 
\author{Andrea Richichi}
\affil{INAF—Osservatorio Astrofisico di Arcetri, Largo E. Fermi 5, I-50125 Firenze, Italy}, 
\author{Puji Irawati}
\affil{National Astronomical Research Institute of Thailand, 191 Siriphanich Bldg., Huay Kaew Rd., Suthep, Muang, Chiang Mai
50200, Thailand}, 
\author{Neelam Panwar}
\affil{Aryabhatta Research Institute of Observational Sciences, Manora Peak, Nainital 263002, Uttarakhand, India},                       
\author{V. S. Dhillon}
\affil{University of Dept of Physics \& Astronomy, University of Sheffield, Sheffield S3 7RH, UK} 
\affil{Instituto de Astrofisica de Canarias, E-38205 La Laguna, Tenerife, Spain}
\author{T. R. Marsh}
\affil{Department of Physics, University of Warwick, Coventry CV4 7AL, UK}


\begin{abstract}
We present time series photometry in the field of Stock 8 and identified 130 short period variable stars.
Twenty eight main-sequence and 23 pre-main-sequence
variables are found to be part of cluster Stock 8. The main-sequence variables
are classified as slow pulsator of the B type, $\beta$ Cep and $\delta$ Scuti stars. Fourteen
main-sequence stars could be new class variables as discussed by Mowlavi et al. (2013)
and Lata et al. (2014). The age and mass of pre-main-sequence variables
are found to be $\lesssim$ 5 Myr and in the mass range of 0.5 to 2.8 M$_{\odot}$, respectively.  
These pre-main-sequence stars could be T-Tauri variables. 
We have found 5 and 2 of 23 PMS variables as
classical T-Tauri stars and Herbig Ae/Be stars, respectively, whereas 16 PMS stars are classified as weak-line T Tauri stars.
\end{abstract}


\section{Introduction}

The study of variable stars is vital to understand the physical properties of stars as well as their circumstellar environment. Therefore, we are pursuing studies of variable stars in young star clusters to study the evolution of pre-main-sequence (PMS) stars. Star clusters are unique laboratories to study the stellar evolution as they provide a sample of stars having approximately the same age, distance, initial composition and spanning range in masses. Additionally, light curves of several stars can be produced simultaneously to identify variable stars among them.
Stars on the main-sequence (MS) as well as on the PMS may show variability in their light curves. The brightness may vary in few hours or it may take years. 

The largest group of PMS stars is the T Tauri stars (TTSs) (Joy 1945). These TTSs are found along the Milky Way in star forming regions embedded in gas and dust.  The TTSs are slowly contracting to the MS and the location of these stars in the H-R diagram is just above the MS. PMS stars are classified as classical T-Tauri stars (CTTSs), weak-line T-Tauri stars (WTTSs) (Menard \& Bertout 1999) and Herbig Ae/Be (HAeBe) stars. CTTSs are stars with evidence of an accretion disc, whereas WTTSs have little or no accretion disc. HAeBe stars have relatively higher-mass ($\gtrsim$3 $M_{\odot}$) in comparison to TTSs.
 Most WTTSs have simple periodic variations, whereas the brightness variations in CTTSs can be complex and irregular (Herbst et al. 1994). 
Bouvier et al. (1997), Grankin et al. (2007, 2008) and Percy et al. (2010) studied variability in PMS stars. Recently, Lata et al. (2012, 2014) presented light curves of several PMS stars. 

Stock 8 ($l$=173.371 deg, $b$=-0.183 deg) is an extremely young stellar open cluster which is located within the HII region IC 417 (Sh2-234) in the Auriga constellation. It contains many OB stars. A detailed study of this cluster to understand the star formation history, PMS population and the initial mass function (IMF) was presented by Jose et al. (2008). They have determined its fundamental properties such as reddening, distance, age and IMF. The reddening $E(B-V)$ towards the cluster was found to vary between 0.40 to 0.60 mag. The cluster is located at a distance of 2.05 kpc and the radial extent of the cluster is found to be $\sim$6 arcmin ($\sim$3.6 pc). They also identified $H{\alpha}$ emission and near-infrared (NIR) excess young stellar objects (YSOs) using $H{\alpha}$ slitless spectroscopy and Two Micron All Sky Survey (2MASS) NIR data, respectively. It is found that the majority of the PMS stars have ages less than 5 Myr, whereas massive stars in the cluster region have an average age of $\le$ 2 Myr.

Jose et al. (2017) presented deep $VI$ optical photometry along with $JHK$ and 3.6 and 4.5 $\micron$ photometry from UKIDSS and Spitzer-IRAC and studied the stellar content and star formation processes in the young cluster Stock 8. The age of the cluster was estimated as 3 Myr with an age spread of $\sim$0.25 Myr. The fraction of YSOs surrounded by disks is found as $\sim$35\%. Jose et al. (2017) also identified several Class I and Class II YSOs within the Stock 8 region  on the basis of color excess in the $J$, $H$, $K$, 3.6, and 4.5 $\micron$ bands.

Stock 8, being an extremely young cluster, is an interesting object which contains a number of PMS, O/B type as well as other MS stars.
Since the population of variable stars in the region of Stock 8 has not been 
studied till now, we have carried out time series photometry of the region containing  Stock 8 to search for the variable
stars. The time series observations of Stock 8 have been taken on 18 nights to find short period variables within the cluster. 
 The observations, procedure of data reduction, identification of variables and determination of period are presented in Section 2. Section 3 deals with the association of variable stars detected in the present work with the cluster Stock 8 on the basis of two colour diagram (TCD) and colour-magnitude diagram (CMD).  The estimation of the mass and age of YSOs is described in Section 4. In Section 5, we study the spectral energy distribution of identified YSOs. Identified variables are characterized in Section 6. In section 7, the effect of NIR excess on the rotation period was studied. Section 8 summarizes results obtained in the present study.

\begin{table}
\centering
\caption{Log of optical photometric CCD observations. `N' and `Exp.' refer to number of frames  and exposure time respectively. \label{tab:obsLog}}
\begin{tabular}{lccc}
\hline
Date& Filter & (N$\times$ Exp.) & Telescope \\
& & \\
\hline
12 January 2015 &  $g^{\prime}$ &268$\times$13s&2.4 m TNO \\
13 January 2015 &  $g^{\prime}$ &538$\times$13s&2.4 m TNO \\
14 January 2015 &  $g^{\prime}$ &342$\times$13s&2.4 m  TNO\\
08 March 2015   &  $g^{\prime}$ &808$\times$13s&2.4 m  TNO\\
10 March 2015   &  $g^{\prime}$ &694$\times$13s&2.4 m  TNO\\
06 January 2016 &  $g^{\prime}$ &214$\times$13s&2.4 m  TNO\\
10 January 2016 &  $g^{\prime}$ &111$\times$13s&2.4 m  TNO\\
01 March 2016   & $V$ & 83$\times$15s&2.3 HCT \\
02 March 2016   & $V$ & 100$\times$13s&2.3  HCT \\
15 November 2015 & $V$ & 56$\times$150s& 0.5 m TNO\\
17 November 2015 & $V$ & 46$\times$90s& 0.5 m TNO\\
19 November 2015 & $V$ & 47$\times$90s& 0.5 m TNO\\
23 November 2015 & $V$ & 09$\times$90s& 0.5 m TNO\\
28 November 2015 & $V$ & 205$\times$90s& 0.5 m TNO\\
29 November 2015 & $V$ & 84$\times$90s& 0.5 m TNO\\
09 December 2015 & $V$ & 163$\times$90s& 0.5 m TNO\\
23 December 2014 & $V$ & 100$\times$60s& 1.04 m ARIES\\
15 January  2015 & $V$ & 49$\times$60s& 1.04 m ARIES\\
\hline
\end{tabular}
\end{table}
\begin{figure*}
\hbox{
\hspace{-2 cm}
\includegraphics[width=17cm]{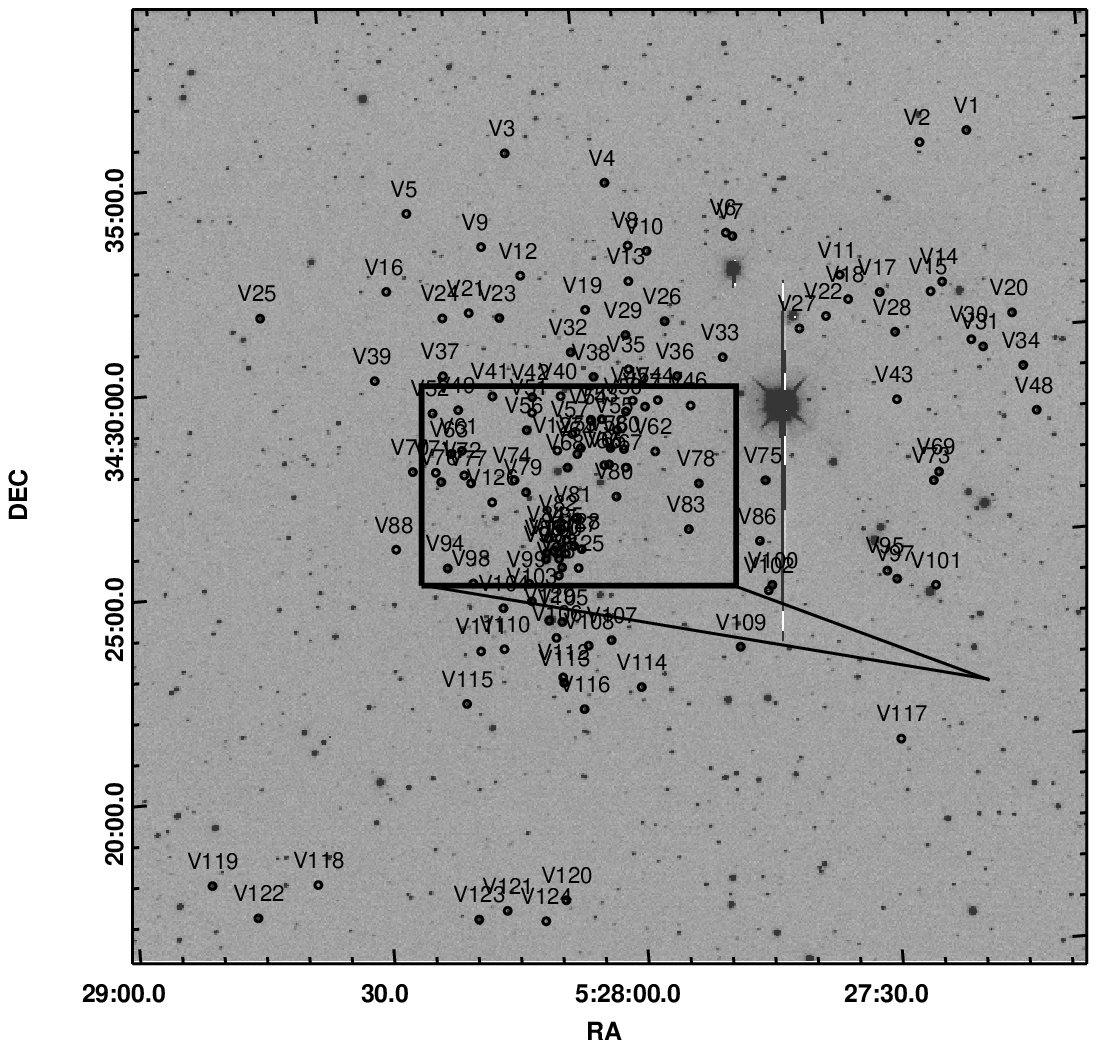}
\hspace{-5.5 cm}
\includegraphics[width=6cm]{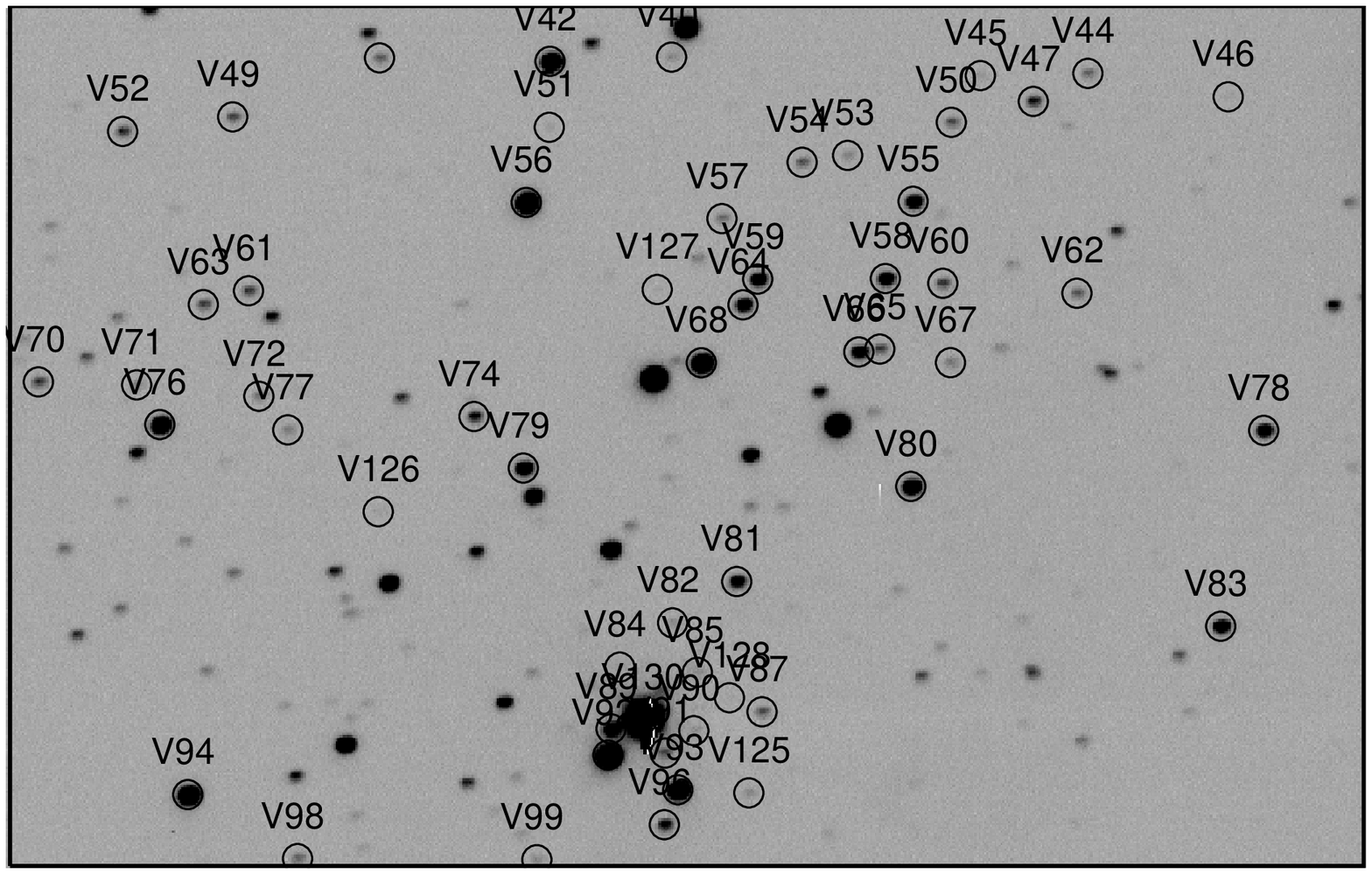}
}
\caption{ The observed region of Stock 8 in $V$ band. Encircled and labeled with numbers are variable candidates detected in the present study. The equinox J2000.0 is used for equatorial coordinates RA and DEC. }
\end{figure*}

\begin{figure}
\includegraphics[width=8cm]{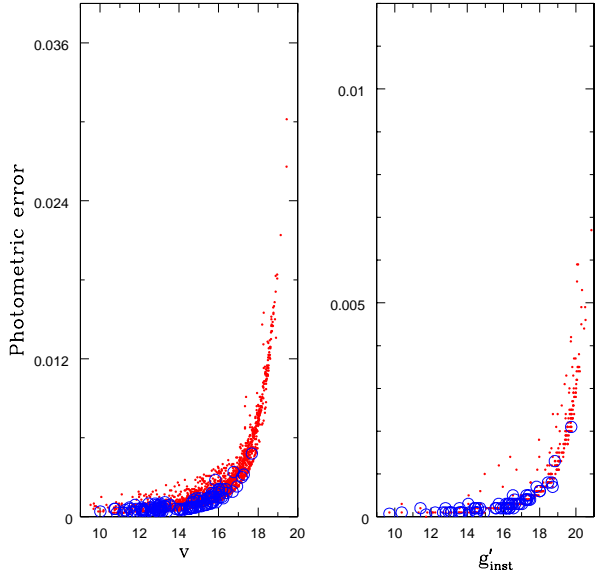}
\caption{Standard photometric error of all the detected stars. The open circles represent variables identified in the present work. In $V$ band we detected 126 variables, whereas in $g^{\prime}$ band we detected 64 variables. The total number of detected variables in both bands is 130.
}
\end{figure}

\begin{figure*}
\vbox{
\includegraphics[width=16.cm, height=13cm]{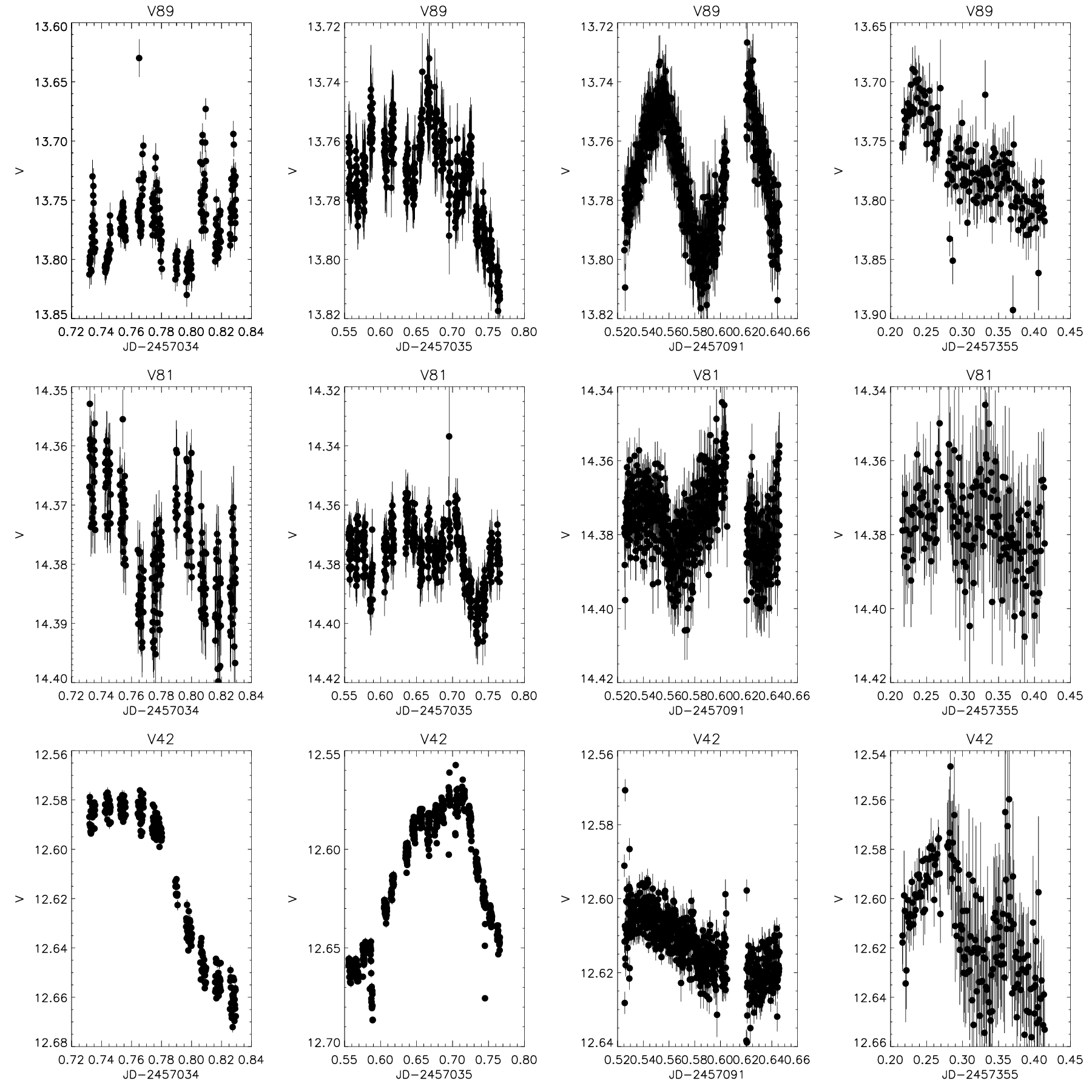}
\includegraphics[width=16.cm, height=13cm]{var_real1.ps}
}
\vspace{-5cm}
\caption{The $V$ band sample light curves of a few variable stars identified in the
present work. $V$ represents the standard magnitude in $V$ band.}
\end{figure*}

\section{Data and identification of variables}
\subsection{Observations and Data Reduction}
Photometric observations of young open cluster Stock 8  
were carried out on a total of 18 nights during 2014 to 2016 from four telescopes.
The observations were taken using the 2.4 m Thai National Telescope (TNT) of the
Thai National Observatory (TNO) located on one of the ridges (2457 m) of Doi Inthanon, the highest peak in Thailand. The telescope is 
equipped with ULTRASPEC which has a 1024$\times$1024 pixel$^2$ frame-transfer, electron-multiplying CCD and that together with re-imaging optics provides photometry 
over a field of 7.7$\times$7.7 arcmin$^2$ at frame rates of up to $\sim$200 Hz in window mode (for details see Dhillon et al. 2014; Richichi et al. 2014).  
The observations were taken in $g^{\prime}$ band on 7 nights during 2015 January 12 to 2016 March 10. 

The 0.5 m telescope at TNO was also used for observations of field containing Stock 8. 
An Andor Tech
2048$\times$2048 pixels CCD camera attached to the 0.5 m
Schmidt-Cassegrain Telescope of TNO was used. The plate scale was about 0.684 arcsec/pixel. The resulting field of view of each image
was $\sim$ 23.9$\times$23.9 arcmin$^2$. The observations of Stock 8 were taken in $V$ band from 15 November 2015 to 09 December 2015.

$V$ band photometric imaging of Stock 8 was obtained using the ARIES 1.04 m telescope with a 2k$\times$2k CCD.
The plate scale of 0.37 arcsec/pixel provides a field of view about 13 arcmin $\times$ 13 arcmin.
In order to improve the signal-to-noise ratio (S/N), a 2$\times$2 binning mode was used.

Further, optical observations of Stock 8 region in $V$ band have also been carried out using the 2-m Himalayan Chandra Telescope (HCT) of the Indian Astronomical Observatory (IAO) and 
Himalayan Faint Object
Spectrograph Camera (HFOSC).
The detector used in the HFOSC is a 2k$\times$4k SiTe chip with pixel size 15$\times$15
microns. It has plate scale 0.296
arcsec/pixel which gives 10$\times$10 arcmin$^2$ field of view. 

The observational log is given in Table 1.
In total, 3917 frames in the Stock 8 region were secured on 18 nights. We have taken bias and twilight flat images along with the cluster field. Fig. 1 shows the observed region of Stock 8 open cluster.
The raw CCD images were preprocessed  using Image Reduction and Analysis Facility (IRAF)\footnote{IRAF is distributed by the National Optical Astronomy Observatory, which is operated by the Association of Universities for Research in Astronomy (AURA) under cooperative agreement with the National Science Foundation.}.
The pre-processing of images includes bias subtraction, flat fielding and cosmic
ray removal. We have used DAOPHOT package (Stetson 1987) to determine the instrumental magnitude of the stars. PSF photometry has been carried out to get the instrumental magnitudes of the sources and it 
is mainly used for crowded regions to get better results. 
 Details of the procedure can be found in our earlier papers (Lata et al. 2011, 2012).

In order to find the translation, rotation and scaling solutions between different photometry files, we have used DAOMATCH (Stetson 1992).
DAOMASTER (Stetson 1992) was used to match the point sources.
DAOMASTER uses the output file of DAOMATCH with
transformations and a list of photometry files. It refines the transformations
using all matched stars and derives robust photometric offsets between
frames. We have used  DAOMASTER to remove the effects of frame-to-frame flux variation
due to airmass and exposure time. This task applies an additive constant in order to make the mean flux level for each frame equal to the reference frame. The corrected magnitudes of each frame were listed in a $ .cor$ file.
We have $.cor$ files for $V$ and $g^{\prime}$ bands. The $V$ band $.cor$ contains 1721 sources while $g^{\prime}$ $.cor$ has 361 stars. 
These corrected instrumental magnitudes of stars are further transformed into standard ones in the next section. 

Fig. 2 shows the photometric error of all the stars and identified variables as a function of mean $v$ and $g_{inst}^{\prime}$ instrumental magnitude.
The standard photometric error of the mean magnitude for each star, based on photometric error of individual frame, has been taken from the $.mag$ file given by the DAOMASTER.  
Out of 130 stars, only 126 are plotted in the left panel of Fig. 2. Four stars could not be detected in observations taken in $V$ band.
Right panel of Fig. 2 displays 64 variables which were detected in $g^{\prime}$ band.
 The standard photometric error of mean magnitude in $g^{\prime}$ band is found to be much smaller than that of $V$ band as shown in Fig. 2. 
This could be due to large number of observations in $g^{\prime}$ band compared to those in $V$ band. 

\begin{figure}
\includegraphics[width=9cm]{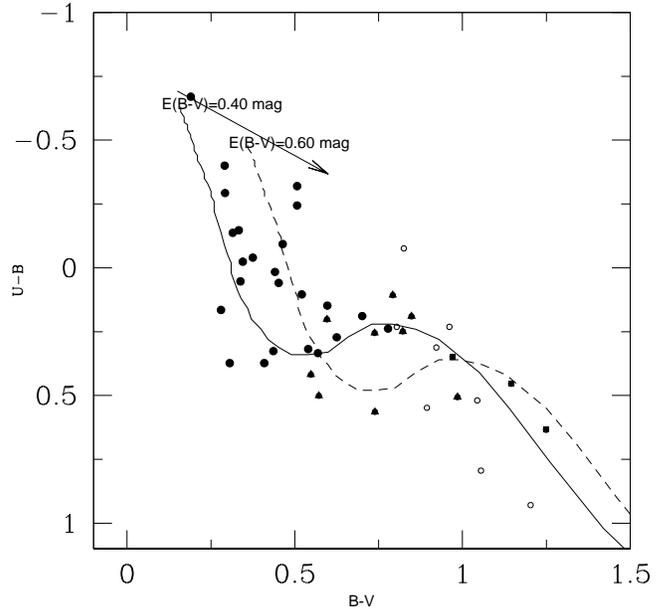}
\caption{$(U-B)/(B-V)$ TCD for variables detected in the present work. The $UBV$ data are taken from 
Jose et al. (2008). The continuous and dashed curve represent the zero-age-main-sequence (ZAMS) (Girardi et al. 2002) which are moved in the direction of the reddening vector for reddening $E(B-V)= 0.4$ mag and 0.60 mag. Filled and open circles represent MS and field/unclassified variables, respectively. Filled squares and triangles show PMS and BP population stars, respectively.  
The straight line indicates reddening vector for $E(U-B)/E(B-V)$ = 0.72.}
\end{figure}

\subsection{Archival data}
The $UBVI$ data of Stock 8 have been taken from Jose et al. (2008, 2017), whereas the NIR data have been taken from the 2MASS Point Source Catalogues (PSC; Cutri et al. 2003). The 2MASS counterparts of the variables were searched using a match value of 3 arcsec. The 2MASS magnitudes and colours were transformed to the Caltech (CIT) photometric system using the relations given on the 2MASS web site \footnote {http://www.astro.caltech.edu/$\sim$jmc/2mass/v3/transformations/}. We have also used NASA's Wide-field Infrared Survey Explorer (WISE) data (Cutri et al. 2012) taken at wavelengths 3.4, 4.6, 12.0, and 22.0 $\micron$.

\begin{table}
\centering
\caption{The sample of data for the variable stars. The complete data of all the variable stars are available in electronic form. }
\begin{tabular}{cccc}
\hline
ID& JD&   V&      $\sigma_{V}$ \\
 & &  (mag)& (mag)     \\
\hline
V1 &2457355.2165970&     12.880&      0.021 \\
V1 &2457355.2187500&     12.874&      0.022 \\
V1 &2457355.2208910&     12.893&      0.019 \\
V1 &2457355.2230440&     12.878&      0.020 \\ 
... &...            &      ...  &      ...   \\
... &...            &      ...  &      ...   \\ 
... &...            &      ...  &      ...   \\
... &...            &      ...  &      ...   \\
\hline
\end{tabular}
\end{table}

\begin{table*}
\caption{The photometric data, period, amplitude, time of maximum brightness $T_{0}$ and classification of 130 variables in the region of Stock 8.
The $UBVI$ and NIR data have been taken from Jose et al. (2008, 2017) and 2MASS point source catalogue (Cutri et al. 2003), respectively.
Column 15 infers classification (c1) of stars as per Jose et al. (2017), while column 16 shows present classification (c2) of variables.
}
\tiny
\begin{tabular}{llllllllllllllll}
\hline
ID  &  RA(2000)&   Dec(2000)&      $V$&   $U-B$&    $B-V$&   $V-I$&      $J$&    $H$&   $K$&    Period& Amp. &$T_{0}$ & prob. & c1 & c2 \\
&         &            &           (mag)& (mag) &   (mag)&  (mag)   &  (mag)&   (mag)&  (mag)&    (days) & (amp)& (2457000.+)    &   &    \\
\hline
V1  &    81.808528 &  34.582028  &           -  &         - &          - &          - &           - &       - &       - &  0.174048  & 0.027  &355.344271    &    0.15     & -              &   unclass.  \\ 
V2  &    81.832167 &  34.578667  &           -  &         - &          - &          - &           - &       - &       - &  0.205270  & 0.093  &355.411887    &          1.     & -              &   unclass.  \\ 
V3  &    82.037417 &  34.587694  &           -  &         - &          - &          - &      11.155 &  10.577 &  10.446 &  0.218461  & 0.034  &355.314306    &          0.     & -              &   Field       \\ 
V4  &    81.989389 &    34.5725  &           -  &         - &          - &          - &           - &       - &       - &  0.232714  & 0.025  &366.241551    &    0.99     & -              &   unclass.  \\ 
V5  &    82.088444 &   34.56625  &      16.583  &         - &      0.998 &      1.125 &      14.702 &  14.275 &  14.188 &  0.262154  & 0.123  &356.201551    &          1.     & -              & BP            \\ 
V6  &    81.931556 &  34.548167  &           -  &         - &          - &          - &           - &       - &       - &  0.250823  & 0.067  &355.218750    &          1.     & -              &   unclass.  \\ 
V7  &      81.9285 &  34.546583  &           -  &         - &          - &          - &           - &       - &       - &  0.341004  & 0.037  &350.175995    &          1.     & -              &   unclass.  \\ 
V8  &    81.980583 &  34.546111  &           -  &         - &          - &          - &           - &       - &       - &  0.363382  & 0.130  &350.175995    &          1.     & -              &   unclass.  \\ 
V9  &    82.052972 &  34.550306  &      16.952  &         - &      0.981 &      1.209 &      14.884 &   14.34 &  14.278 &  0.387657  & 0.158  &356.201551    &          1.     & -              &   Field       \\ 
V10 &      81.9715 &  34.543333  &           -  &         - &          - &          - &           - &       - &       - &  0.232714  & 0.023  &366.241551    &          1.     & -              &   unclass.  \\ 
V11 &    81.877167 &  34.527278  &           -  &         - &          - &          - &           - &       - &       - &  0.352351  & 0.128  &350.175995    &          1.     & -              &   unclass.  \\ 
V12 &    82.034833 &  34.537361  &      16.655  &         - &      1.192 &       1.28 &      14.461 &  13.908 &  13.814 &  0.377979  & 0.116  &015.372340    &          1.     & -              &    Field      \\ 
V13 &    81.981722 &  34.531611  &           -  &         - &          - &          - &           - &       - &       - &  0.203531  & 0.056  &355.218750    &          1.     & -              &   unclass.  \\ 
V14 &    81.826944 &  34.521111  &           -  &         - &          - &          - &           - &       - &       - &  0.372874  & 0.121  &344.233750    &    0.18     & -              &       unclass.  \\ 
V15 &    81.833083 &  34.517556  &           -  &         - &          - &          - &           - &       - &       - &  0.267183  & 0.100  &355.223044    &          1.     & -              &   unclass.  \\ 
V16 &    82.101611 &  34.535111  &       15.82  &      0.19 &      0.848 &      1.001 &      14.167 &  13.784 &  13.724 &  0.377563  & 0.096  &350.177106    &          1.     & -              &  BP           \\ 
V17 &    81.858194 &  34.518917  &           -  &         - &          - &          - &           - &       - &       - &  0.484089  & 0.040  &366.219016    &          1.     & -              &   unclass.  \\ 
V18 &    81.874056 &  34.517083  &           -  &         - &          - &          - &           - &       - &       - &  0.267183  & 0.082  &356.276539    &          1.     & -              &   unclass.  \\ 
V19 &    82.004222 &  34.521333  &       16.64  &         - &      2.036 &      2.326 &      12.604 &  11.737 &  11.426 &  0.377159  & 0.128  &356.204769    &          1.     & -              & PMS(WTTS)           \\ 
V20 &    81.793806 &  34.506194  &           -  &         - &          - &          - &           - &       - &       - &  0.319554  & 0.019  &346.320949    &          0.     & -              &   unclass.  \\ 
V21 &    82.061806 &  34.523806  &      16.568  &         - &      0.858 &      1.112 &      14.716 &  14.431 &  14.225 &  0.376725  & 0.104  &356.205845    &          1.     & -              &  BP           \\ 
V22 &    81.885722 &  34.510944  &           -  &         - &          - &          - &           - &       - &       - &  0.296789  & 0.049  &350.175995    &          1.     & -              &   unclass.  \\ 
V23 &    82.046917 &  34.520806  &      12.892  &     0.318 &       0.54 &      0.661 &      11.743 &   11.56 &   11.49 &  0.170225  & 0.036  &366.219016    &          1.     & -              & MS(new class)            \\ 
V24 &    82.075056 &  34.522472  &      16.607  &         - &      0.928 &      1.088 &      14.786 &  14.349 &  14.193 &  0.377979  & 0.094  &356.205845    &          1.     & -              &  BP           \\ 
V25 &    82.164944 &   34.52825  &      13.878  &     0.929 &      1.203 &      1.258 &      11.481 &  10.907 &  10.751 &  0.179500  & 0.035  &342.389525    &    0.41     & -              &    Field      \\ 
V26 &    81.965528 &  34.514111  &           -  &         - &          - &          - &       9.075 &   8.441 &   8.219 &  0.363838  & 0.082  &366.241551    &          1.     & -              & PMS(WTTS)           \\ 
V27 &    81.899333 &  34.506722  &           -  &         - &          - &          - &           - &       - &       - &  0.228593  & 0.050  &350.177106    &          1.     & -              &   unclass.  \\ 
V28 &    81.852306 &  34.502222  &           -  &         - &          - &          - &           - &       - &       - &  0.295748  & 0.022  &342.393125    &    0.04     & -              &   unclass.  \\ 
V29 &    81.985389 &  34.509778  &      13.333  &         - &          - &      0.902 &      11.846 &  11.581 &  11.477 &  0.260328  & 0.037  &355.318600    &          1.     & -              & MS            \\ 
V30 &    81.815028 &  34.496639  &           -  &         - &          - &          - &           - &       - &       - &  0.210166  & 0.109  &356.271169    &          1.     & -              &    Field      \\ 
V31 &      81.8095 &  34.493333  &           -  &         - &          - &          - &           - &       - &       - &  0.112561  & 0.042  &350.177106    &           1.    & -              &    $\delta$ Scuti? \\ 
V32 &    82.013194 &  34.504528  &      13.363  &         - &      6.442 &      3.604 &       7.124 &   5.671 &    4.75 &  0.352105  & 0.040  &355.362512    &          1.     & -              &    Field      \\ 
V33 &    81.938389 &  34.497556  &           -  &         - &          - &          - &           - &       - &       - &  0.420986  & 0.099  &344.198507    &          1.     & -              &   unclass.  \\ 
V34 &    81.790556 &  34.484417  &           -  &         - &          - &          - &           - &       - &       - &  0.281200  & 0.022  &366.266817    &    0.01     & -              &   unclass.  \\ 
V35 &    81.985278 &  34.495722  &      16.685  &         - &      1.254 &      1.524 &        14.1 &  13.525 &  13.308 &  0.144190  & 0.065  &034.828316    &          1.     & -              &    Field      \\ 
V36 &    81.961611 &  34.491361  &           -  &         - &          - &          - &      12.109 &  11.948 &   11.79 &  0.127839  & 0.092  &036.828517    &          1.     & -              &    Field      \\ 
V37 &    82.077139 &  34.498917  &      16.368  &         - &      1.788 &      2.426 &      12.254 &  11.417 &  11.148 &  0.427844  & 0.093  &449.265729    &          1.     & -              & PMS(WTTS)           \\ 
V38 &    82.002917 &  34.493806  &      12.629  &     0.148 &      0.597 &      0.643 &      11.287 &  11.013 &  10.954 &  0.052942  & 0.024  &089.636788    &          1.     & -              &    MS(SPB)      \\ 
V39 &    82.110972 &  34.499111  &      15.385  &     0.249 &      0.822 &       1.06 &      13.605 &  13.202 &  13.121 &  0.267509  & 0.034  &350.175995    &          1.     & -              &  BP           \\ 
V40 &    82.019944 &  34.486944  &      16.667  &         - &      0.973 &      1.059 &      14.984 &  14.587 &  14.518 &  0.307807  & 0.097  &038.167660    &          1.     & -              &  BP           \\ 
V41 &    82.053556 &  34.489083  &      15.878  &         - &      0.837 &      1.148 &      13.828 &  13.427 &  13.291 &  0.182051  & 0.018  &036.830146    &          1.     & -              &  BP           \\ 
V42 &    82.033972 &  34.487472  &      12.665  &    -0.293 &      0.292 &      0.479 &      11.859 &  11.744 &  11.626 &  0.189434  & 0.031  &036.829257    &          1.     & -              & MS(SPB)            \\ 
V43 &     81.85425 &  34.474667  &           -  &         - &          - &          - &           - &       - &       - &  0.180579  & 0.060  &355.230556    &          1.     & -              &   unclass.  \\ 
V44 &    81.972194 &  34.482222  &      16.084  &         - &      0.891 &      1.173 &      14.109 &  13.697 &  13.662 &  0.230325  & 0.059  &089.632938    &          1.     & -              &  BP           \\ 
V45 &    81.984583 &  34.482833  &      17.323  &         - &      1.733 &      2.281 &      13.383 &  12.536 &  12.209 &  0.498848  & 0.239  &038.339120    &          1.     & -              & PMS(WTTS)           \\ 
V46 &    81.956278 &  34.478917  &           -  &         - &          - &          - &           - &       - &       - &  0.549653  & 0.408  &015.281560    &          1.     & -              &   unclass.  \\ 
V47 &     81.97875 &  34.479972  &      15.143  &     0.107 &      0.792 &      0.987 &      13.469 &   13.14 &  13.037 &  0.144650  & 0.060  &036.828517    &          1.     & -              &  BP           \\ 
V48 &    81.785833 &  34.465722  &           -  &         - &          - &          - &           - &       - &       - &  0.220069  & 0.026  &355.314306    &          0.     & -              &   unclass.  \\ 
V49 &    82.071028 &  34.484583  &      15.605  &     0.519 &      1.044 &      1.346 &      13.275 &  12.823 &  12.666 &  0.128532  & 0.024  &091.620475    &          1.     & -              &    Field      \\ 
V50 &    81.988389 &  34.478611  &      15.715  &     0.231 &      0.804 &      0.906 &       14.04 &   13.75 &  13.612 &  0.480021  & 0.047  &091.645200    &          1.     & -              &    Field      \\ 
V51 &    82.034694 &  34.481194  &      17.773  &         - &      1.388 &      1.668 &      14.861 &  14.177 &   13.91 &  0.168994  & 0.041  &091.638094    &          1.     & -              & PMS(WTTS)           \\ 
V52 &    82.083833 &  34.484028  &      15.416  &     0.501 &      0.572 &      0.921 &      13.789 &  13.574 &  13.415 &  0.114324  & 0.026  &091.620475    &          1.     & -              &  BP           \\ 
V53 &    82.000639 &   34.47625  &      16.679  &         - &       0.98 &      1.149 &      14.761 &  14.309 &  14.227 &  0.334858  & 0.047  &036.820481    &          1.     & -              &  BP           \\ 
V54 &    82.005944 &  34.475944  &      15.825  &     0.564 &      0.739 &      0.986 &      14.098 &  13.775 &   13.62 &  0.234592  & 0.022  &036.831182    &          1.     & -              &  BP           \\ 
V55 &    81.993556 &  34.471389  &      13.934  &     0.272 &      0.625 &      0.795 &      12.577 &  12.323 &  12.246 &  0.120563  & 0.035  &036.828517    &          1.     & -              & MS(new class)            \\ 
V56 &    82.038083 &  34.474222  &      11.888  &      -0.4 &      0.291 &      0.479 &      11.069 &  10.948 &  10.907 &  0.252462  & 0.019  &355.362512    &          1.     & III/field  & MS(SPB)            \\ 
V57 &    82.015722 &  34.471194  &      16.077  &         - &      1.075 &      1.409 &       13.55 &  13.181 &   12.95 &  0.224849  & 0.017  &036.820629    &    0.99     & III/field  &    Field      \\ 
V58 &      81.9975 &   34.46425  &      13.978  &     0.794 &      1.055 &      1.256 &       11.87 &   11.29 &  11.178 &  0.141626  & 0.028  &036.828517    &          1.     & III/field  &    Field      \\ 
V59 &    82.012194 &  34.465167  &       14.08  &    -0.093 &      0.464 &      0.702 &      12.834 &  12.592 &  12.479 &  0.364331  & 0.014  &036.829405    &          1.     & III/field  & MS(new class)            \\ 
V60 &    81.990944 &  34.463361  &      15.491  &     0.312 &      0.922 &      1.078 &      13.647 &  13.243 &  13.147 &  0.166027  & 0.023  &089.635751    &    0.93     & III/field  &    Field      \\ 
V61 &    82.070889 &  34.467917  &      15.565  &     0.634 &      1.249 &      1.568 &      12.755 &   11.93 &  11.433 &  0.128722  & 0.026  &089.560241    &          1.     & -              & PMS(CTTS)           \\ 
V62 &    81.975611 &  34.461389  &       15.82  &     0.506 &      0.985 &      1.095 &      13.884 &  13.463 &  13.334 &  0.145872  & 0.046  &036.820481    &    0.95     & -              &  BP           \\ 
V63 &    82.076222 &  34.466944  &       15.66  &     0.255 &      0.738 &      1.075 &       13.93 &  13.529 &  13.444 &  0.113297  & 0.025  &089.558021    &    0.85     & -              &  BP           \\ 
V64 &    82.014139 &  34.462833  &      14.042  &     0.059 &      0.452 &      0.662 &      12.784 &  12.493 &  12.363 &  0.196151  & 0.010  &366.241551    &          1.     & III/field  & MS(new class)            \\ 
V65 &    81.998833 &  34.457583  &      15.719  &     0.202 &      0.596 &      1.026 &      14.103 &  13.756 &  13.702 &  0.229548  & 0.011  &355.218750    &          1.     & III/field  &  BP           \\ 
V66 &    82.001278 &  34.457472  &      14.178  &     0.373 &      0.409 &      0.491 &      13.292 &  13.174 &  13.095 &  0.094072  & 0.014  &089.531074    &          1.     & III/field  & MS($\delta$ Scuti)            \\ 
V67 &    81.990833 &   34.45575  &      16.418  &         - &      0.926 &      1.098 &      14.567 &  14.106 &  14.006 &  0.180390  & 0.014  &366.259306    &          1.     & III/field  &  BP           \\ 
V68 &      82.0195 &  34.457639  &      12.229  &     0.189 &      0.701 &      0.862 &      10.774 &  10.436 &  10.322 &  0.216378  & 0.040  &035.754519    &          1.     & III/field  & MS(SPB)            \\ 
V69 &    81.836583 &  34.443833  &           -  &         - &          - &          - &           - &       - &       - &  0.273521  & 0.023  &366.293657    &    0.99    & -              &   unclass.  \\ 
V70 &    82.095944 &  34.460833  &      15.495  &     0.231 &      0.961 &      1.151 &      13.505 &  13.099 &   12.94 &  0.114027  & 0.034  &450.124711    &          1.     & -              &    Field      \\ 
V71 &    82.084667 &  34.459806  &      16.618  &         - &      0.866 &       1.09 &      14.826 &  14.374 &  14.259 &  0.093118  & 0.026  &366.211505    &          1.     & -              &  BP           \\ 
V72 &    82.070722 &  34.457806  &      16.286  &         - &      0.813 &      1.035 &      14.573 &  14.218 &  14.033 &  0.120263  & 0.024  &089.559353    &          1.     & III/field  &  BP           \\ 
V73 &    81.839556 &  34.440472  &           -  &         - &          - &          - &           - &       - &       - &  0.254942  & 0.022  &366.241551    &          1.     & -              &   unclass.  \\ 
V74 &    82.046167 &   34.45425  &      14.961  &         - &      1.632 &      1.782 &      11.832 &  11.065 &  10.834 &  0.147866  & 0.030  &089.559797    &    0.01     & III/field  & PMS(WTTS)           \\ 
V75 &    81.922556 &  34.446028  &           -  &         - &          - &          - &       11.06 &  10.958 &  10.903 &  0.267183  & 0.066  &355.364664    &          1.     & -              &    Field      \\ 
V76 &    82.082389 &  34.455861  &      12.806  &     0.373 &      0.306 &       0.41 &      12.078 &  11.987 &  11.942 &  0.111308  & 0.025  &089.559353    &          1.     & -              & MS($\delta$ Scuti)            \\ 
V77 &    82.067722 &  34.454389  &      16.281  &         - &      1.904 &      2.189 &      12.445 &  11.541 &  11.249 &  0.249165  & 0.036  &089.637232    &    0.91     & III/field  & PMS(WTTS)           \\ 
V78 &    81.955472 &  34.446917  &           -  &         - &          - &          - &      12.303 &  11.935 &   11.82 &  0.209675  & 0.022  &091.568654    &    0.31     & -              &    Field      \\ 
\hline                                                                                                                                                        
\end{tabular}                                                                                                                                                 
\end{table*}                                                                                                                                                  
\setcounter{table}{2}                                                                                                                                         
\begin{table*}
\caption{Continued
}
\tiny
\begin{tabular}{llllllllllllllll}
\hline
ID  &  RA(2000)&   Dec(2000)&      $V$&   $U-B$&    $B-V$&   $V-I$&      $J$&    $H$&   $K$&    Period& Amp. &$T_{0}$ & prob. & c1 & c2 \\
&         &            &           (mag)& (mag) &   (mag)&  (mag)   &  (mag)&   (mag)&  (mag)&    (days) & (amp)& (2457000.+)     &   &    \\
\hline
V79 &       82.041 &  34.448972  &      13.717  &     0.053 &      0.338 &      0.512 &      12.816 &  12.715 &  12.621 & 0.183882  & 0.028 &089.545288  &         1.   & III/field  & MS(new class)            \\
V80 &    81.996611 &  34.444278  &      13.211  &     0.165 &       0.28 &      0.411 &      12.459 &  12.283 &   12.26 & 0.128475  & 0.021 &036.828665  &         1.   & III/field  & MS(new class)            \\
V81 &    82.017556 &  34.436556  &      14.397  &     0.334 &      0.569 &       0.72 &      13.219 &  13.002 &  12.774 & 0.063815  & 0.010 &089.559945  &         1.   & III/field  & MS(new class)            \\
V82 &    82.025333 &  34.433111  &      17.244  &    -      &      1.103 &       1.62 &      14.352 &  13.732 &  13.503 & 0.171297  & 0.047 &355.407593  &         1.   & III/field  & PMS(WTTS)           \\
V83 &    81.962333 &  34.428611  &      13.963  &    -0.076 &      0.825 &       0.92 &      12.329 &  12.098 &  11.981 & 0.123067  & 0.042 &089.631309  &   0.0   & -              &    Field      \\
V84 &    82.031833 &  34.429306  &      17.633  &         - &          - &      1.711 &      14.566 &    13.9 &  13.602 & 0.352718  & 0.304 &342.382350  &         1.   & III/field  & PMS(WTTS)           \\
V85 &    82.022972 &  34.428222  &       16.91  &         - &      1.138 &      1.557 &      14.414 &  13.761 &  13.522 & 0.144094  & 0.075 &089.530482  &         1.   & III/field  & PMS(WTTS)           \\
V86 &    81.927806 &  34.421528  &           -  &         - &          - &          - &      12.303 &  12.011 &  11.916 & 0.377563  & 0.026 &366.241551  &         1.   & -              &    Field      \\
V87 &    82.015917 &  34.423972  &      15.715  &     0.454 &      1.145 &      1.421 &      13.233 &  12.639 &  12.422 & 0.306247  & 0.023 &036.828665  &         1.   & III/field  & PMS(WTTS)           \\
V88 &    82.107361 &   34.42975  &      18.788  &         - &      1.219 &      1.995 &      15.236 &  14.771 &  14.388 & 0.250687  & 0.429 &342.441620  &         1.   & -              &    Field      \\
V89 &    82.033444 &    34.4235  &      13.891  &         - &          - &      1.011 &      12.162 &  11.854 &  11.717 & 0.064978  & 0.018 &036.723177  &         1.   & III/field  &    Field      \\
V90 &    82.023944 &  34.422694  &       16.85  &         - &          - &       1.74 &      13.949 &  13.237 &  12.975 & 0.095051  & 0.157 &355.400081  &         1.   & III/field  & PMS(WTTS)           \\
V91 &    82.027472 &  34.420778  &      15.901  &         - &          - &       1.61 &      13.325 &  12.551 &  12.414 & 0.164707  & 0.072 &089.531074  &    0.21   & III/field  & PMS(WTTS)           \\
V92 &    82.034056 &  34.421028  &      11.048  &         - &          - &      0.325 &      10.514 &  10.469 &  10.449 & 0.173359  & 0.029 &035.648077  &         1.   & III/field  & MS            \\
V93 &    82.026361 &  34.417194  &      12.326  &     0.326 &      0.436 &      0.628 &      11.122 &  10.807 &  10.703 & 0.198500  & 0.024 &089.549877  &         1.   & III/field  & MS(new class)            \\
V94 &     82.08275 &  34.420444  &       12.19  &    -0.244 &      0.507 &      0.754 &      10.876 &  10.689 &  10.606 & 0.184804  & 0.020 &355.364664  &         1.   & III/field  & MS(SPB)            \\
V95 &     81.86625 &  34.405194  &           -  &         - &          - &          - &           - &       - &       - & 0.392364  & 0.020 &342.393125  &    0.99  & -              &   unclass.  \\
V96 &     82.02825 &  34.413972  &        14.8  &     0.349 &       0.97 &      1.285 &       12.44 &  11.693 &  11.055 & 0.158525  & 0.050 &089.560537  &         1.   & III/field  & PMS(HAeBe)           \\
V97 &     81.86175 &  34.401583  &           -  &         - &          - &          - &           - &       - &       - & 0.396248  & 0.038 &366.207211  &         1.   & -              &   unclass.  \\
V98 &     82.07075 &  34.413528  &      15.935  &         - &      1.051 &      1.341 &      13.545 &  12.897 &  12.431 & 0.120824  & 0.042 &089.559353  &         1.   & II         & PMS(HAeBe)           \\
V99 &     82.04325 &  34.411611  &      16.825  &         - &       1.52 &      1.892 &      13.519 &  12.569 &  12.042 & 0.179223  & 0.055 &089.636343  &         1.   & III/field  & PMS(CTTS)           \\
V100&      81.9235 &  34.403194  &           -  &         - &          - &          - &      11.629 &  11.349 &  11.245 & 0.220333  & 0.018 &366.326933  &         1.   & -              &     Field     \\
V101&    81.842917 &  34.397806  &           -  &         - &          - &          - &           - &       - &       - & 0.352334  & 0.254 &342.412882  &         1.   & -              &    unclass. \\
V102&      81.9255 &  34.401083  &      14.168  &         - &          - &      0.972 &      12.567 &  12.293 &  12.212 & 0.272409  & 0.024 &346.336366  &         1.   & -              &  MS           \\
V103&    82.042639 &  34.404361  &      13.124  &     0.104 &      0.521 &      0.776 &       11.75 &  11.537 &  11.399 & 0.279487  & 0.040 &089.589409  &   0.51   & III/field  &  MS(new class)           \\
V104&    82.056917 &  34.402444  &      14.984  &     0.548 &      0.894 &      1.204 &      12.863 &  12.457 &  12.262 & 0.112132  & 0.013 &346.331956  &         0.   & III/field  &     Field     \\
V105&    82.028472 &  34.394944  &      13.361  &    -0.137 &      0.315 &      0.508 &      12.523 &  12.426 &  12.362 & 0.161072  & 0.028 &089.559353  &         1.   & III/field  &  MS(new class)           \\
V106&    82.032056 &  34.388639  &      17.108  &         - &      1.315 &      1.681 &      14.256 &  13.611 &  13.308 & 0.229596  & 0.038 &355.221968  &         1.   & III/field  &  PMS(CTTS)          \\
V107&       82.005 &  34.385944  &      13.291  &    -0.024 &      0.345 &      0.482 &      12.333 &  12.204 &  12.158 & 0.069255  & 0.008 &089.612358  &         1.   & III/field  &  MS(new class)           \\
V108&    82.016528 &  34.384417  &      15.183  &     0.418 &      0.548 &      0.721 &       13.82 &  13.508 &  13.352 & 0.199593  & 0.012 &036.811076  &         0.   & III/field  &   BP          \\
V109&     81.94175 &  34.379028  &      12.427  &     -0.32 &      0.507 &      0.732 &      11.147 &  10.993 &  10.923 & 0.195896  & 0.020 &355.294988  &         1.   & -              &  MS(SPB)           \\
V110&    82.058139 &  34.385722  &      18.627  &         - &          - &      1.828 &      15.351 &  14.699 &  14.396 & 0.648860  & 0.412 &356.251852  &         1.   & III/field  &  PMS(CTTS)          \\
V111&    82.069722 &  34.385583  &      16.519  &         - &      1.153 &      1.246 &      14.419 &  13.856 &  13.745 & 0.102806  & 0.034 &036.821073  &         1.   & -                &     Field     \\
V112&    82.030417 &  34.372306  &      16.756  &         - &          - &      1.195 &      14.475 &  14.066 &  13.886 & 0.171350  & 0.047 &036.829405  &         1.   & -                &   BP          \\
V113&    82.030139 &  34.370111  &      13.922  &         - &      1.957 &      2.189 &       10.05 &     9.1 &    8.81 & 0.130534  & 0.013 &355.294988  &         0.   & -                &  PMS(WTTS)          \\
V114&    81.992139 &  34.365833  &      13.629  &     -0.04 &      0.375 &      0.556 &      12.675 &   12.55 &  12.494 & 0.176857  & 0.015 &366.294722  &         1.   & -                &  MS(new class)           \\
V115&    82.078833 &  34.364639  &      13.705  &         - &          - &      1.144 &      11.708 &  11.232 &  11.111 & 0.275857  & 0.031 &038.347990  &         0.   & -                &     Field     \\
V116&    82.021194 &  34.358694  &      13.329  &    -0.147 &      0.333 &      0.506 &      12.429 &  12.346 &  12.241 & 0.450141  & 0.039 &355.294988  &         1.   & -              &  MS(SPB)           \\
V117&      81.8665 &  34.336306  &           -  &         - &          - &          - &      15.142 &  14.617 &  14.353 & 0.187340  & 0.210 &356.273322  &         1.   & -              &     Field     \\
V118&      82.1595 &  34.295611  &           -  &         - &          - &          - &      14.558 &  14.155 &   13.98 & 0.053956  & 0.068 &355.357141  &         1.   & -              &     Field     \\
V119&    82.211611 &  34.298583  &           -  &         - &          - &          - &      11.545 &  11.381 &  11.313 & 0.050593  & 0.028 &356.214433  &         1.   & -              &     Field     \\
V120&    82.038111 &  34.281444  &      13.214  &     0.016 &      0.441 &      0.645 &      12.089 &  11.948 &  11.834 & 0.140128  & 0.015 &355.372176  &         0.   & -              &  MS(new class)           \\
V121&    82.067417 &  34.279028  &      14.276  &     0.238 &      0.778 &      0.924 &      12.769 &  12.475 &  12.368 & 0.140456  & 0.027 &346.331956  &         0.   & -              &  MS(new class)           \\
V122&    82.190361 &  34.283972  &           -  &         - &          - &          - &      12.053 &  11.883 &  11.819 & 0.249029  & 0.043 &356.228380  &         1.   & -              &     Field     \\
V123&    82.081778 &  34.276417  &           -  &         - &          - &          - &      11.736 &  11.602 &  11.551 & 0.311387  & 0.049 &356.216574  &         1.   & -              &     Field     \\
V124&    82.048889 &  34.273556  &      15.861  &         - &       1.15 &       1.19 &      13.857 &  13.361 &  13.217 & 0.295529  & 0.039 &355.375394  &   0.97   & -              &   BP          \\
V125&    82.018222 &  34.416361  &      17.827  &         - &      1.389 &      1.779 &      14.975 &  13.853 &  13.131 & 0.501000  & 0.360 &035.695247  &          1. & II         &  PMS(CTTS)          \\
V126&    82.058111 &  34.445917  &      17.866  &         - &      1.069 &      1.421 &      15.585 &  14.954 &    14.7 & 0.452000  & 0.116 &036.821073  &          1. & III/field  &  PMS(WTTS)          \\
V127&    82.023861 &  34.464944  &      18.877  &         - &          - &      1.641 &      16.052 &  15.706 &  15.065 & 0.446000  & 0.195 &091.643424  &          1. & III/field  &     Field     \\
V128&    82.019528 &  34.425556  &      18.441  &         - &          - &      1.911 &      15.499 &  14.653 &  14.422 & 0.315000  & 0.456 &089.530926  &          1. & unknown              &  PMS(WTTS) \\
V129&    82.034903 &  34.395958  &      10.149  &         - &          - &      0.387 &       9.317 &   9.256 &   9.231 & 0.229000  & 0.164 &089.632198  &          1. & unknown              &  MS  \\
V130&    82.029761 &  34.424144  &       8.953  &     -0.67 &       0.19 &      0.282 &       8.378 &   8.358 &   8.324 & 0.169000  & 0.381 &036.715420  &          1. & unknown              &  MS($\beta$ Cep)  \\
\hline
\end{tabular}
\end{table*}

\begin{table}
\scriptsize
\caption{Mass and age of probable PMS stars.}

\begin{tabular}{ccccc}
\hline

ID &    Mass(CMD)&    Age(CMD) & Mass(SED)&    Age(SED) \\
&    ($M_{\odot}$) &   (Myr)& ($M_{\odot}$) &(Myr)    \\
\hline
V19&    0.69$\pm0.02$&   $<=$0.10$\pm$0.01&   -                &    -              \\
V26&    -            &   -                &   -                &    -              \\
V37&    0.75$\pm0.02$&   $<=$0.10$\pm$0.01&   -                &    -              \\
V45&    0.55$\pm0.03$&   0.16$ \pm$0.04&      -                &    -              \\
V51&    1.32$\pm0.03$&   4.36$ \pm$0.67&     1.23 $\pm$   0.50 &  4.21$ \pm$  2.30  \\
V61&    2.01$\pm0.10$&   0.64$ \pm$0.10&     2.71 $\pm$   1.00 &  3.21$ \pm$  2.10 \\
V74&    1.56$\pm0.08$&   0.20$ \pm$0.02&      3.67 $\pm$   0.14 &  0.93$ \pm$  1.42 \\
V77&    0.79$\pm0.02$&   $<=$0.10$\pm$0.01&  1.01 $\pm$   0.41 &  1.22$ \pm$  0.35  \\
V82&    1.44$\pm0.05$&   2.64$ \pm$0.44&     2.45 $\pm$   0.60 &  6.35$ \pm$  2.31 \\
V84&    1.25$\pm0.06$&   2.72$ \pm$0.49&      -                &    -              \\
V85&    1.63$\pm0.04$&   2.54$ \pm$0.34&      -                &    -              \\
V87&    2.32$\pm0.05$&   1.49$ \pm$0.31&     1.88 $\pm$   0.52 &  1.85$ \pm$  1.74 \\
V90&    1.20$\pm0.07$&   0.92$ \pm$0.14&     -                  &   -               \\
V91&    1.73$\pm0.09$&   0.69$ \pm$0.09&     -                 &    -              \\
V96&    2.75$\pm0.15$&   1.92$ \pm$0.46&     5.68 $\pm$   0.51 &  4.15$ \pm$  1.43 \\
V98&    2.13$\pm0.08$&   3.15$ \pm$0.67&     3.41 $\pm$   0.29 &  6.39$ \pm$  2.71 \\
V99&    0.93$\pm0.05$&   0.52$ \pm$0.05&     2.47 $\pm$   1.04 &  3.74$ \pm$  2.20 \\
V106&   1.33$\pm0.07$&   1.64$ \pm$0.26&     1.95 $\pm$   0.90 &  4.13$ \pm$  3.30 \\
*V110&   1.04$\pm0.04$&   5.82$ \pm$1.01&     -                &    -              \\
V113&   2.20$\pm0.02$&   0.01$ \pm$0.01&      -                &    -              \\
V125&   1.11$\pm0.06$&   2.43$ \pm$0.41&     2.59 $\pm$   1.37 &  0.49$ \pm$  0.98 \\
V126&   1.29$\pm0.01$&    $>$10    - &       1.83 $\pm$   0.35 &  0.28$ \pm$  0.12 \\
V128&   0.91$\pm0.05$&   2.85$ \pm$0.48&      -                &    -              \\
\hline
\end{tabular}
\end{table}

\begin{figure}
\includegraphics[width=9cm]{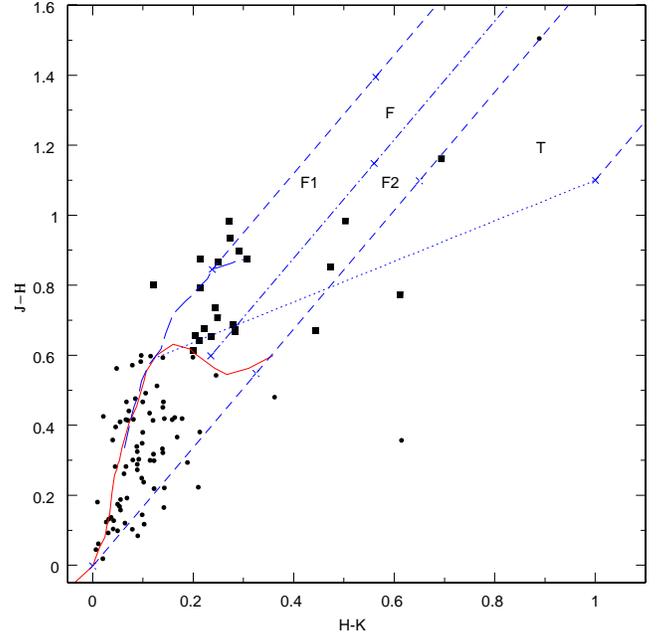}
\caption{$(J-H)/(H-K)$ TCD for variables identified in the field of
 Stock 8. $JHK$ data are taken from the 2MASS catalogue (Cutri et al. 2003). The continuous and long dashed curve show sequences for dwarfs and giants (Bessell \& Brett 1988), respectively. The locus of TTSs (Meyer et al. 1997) is represented by dotted curve.
The reddening vectors (Cohen et al. 1981) are shown by small dashed lines and
an increment of visual extinction of $A_{V}$ = 5 mag is denoted by crosses on the reddening vectors. 
 The `F' region and `T' region are mentioned in Section 3, where the sub-regions
`F1' and `F2' are discussed  in Section 6.2. Filled squares and circles represent PMS and other (MS, BP, field/unclassified) variables, respectively.} 

\end{figure}

\begin{figure}
\includegraphics[width=9cm]{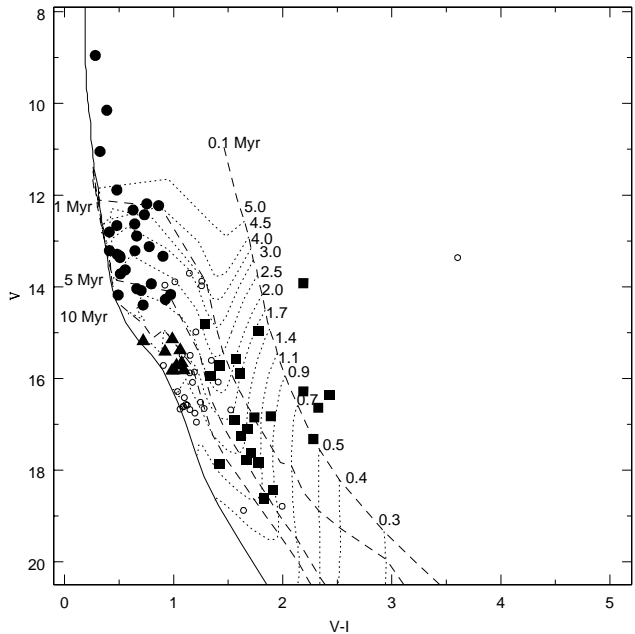}
\caption{$V/(V-I)$ CMD for variable stars detected in the cluster Stock 8. The $V$ and $I$ data are taken from Jose et al. (2008).
The filled squares represent probable TTSs, whereas filled circles, open circles and filled triangles show MS stars, field/unclassified and BP population in the direction of the cluster, respectively.
The continuous curve shows ZAMS by Girardi et al. (2002) while dashed lines
 represent PMS isochrones for 
0.1, 1, 5, 10 Myrs (Siess et al. 2000).
The PMS evolutionary tracks of stars for different masses taken from Siess et al. (2000) are shown by dotted curves.
}
\end{figure}

\begin{figure}
\includegraphics[width=9cm]{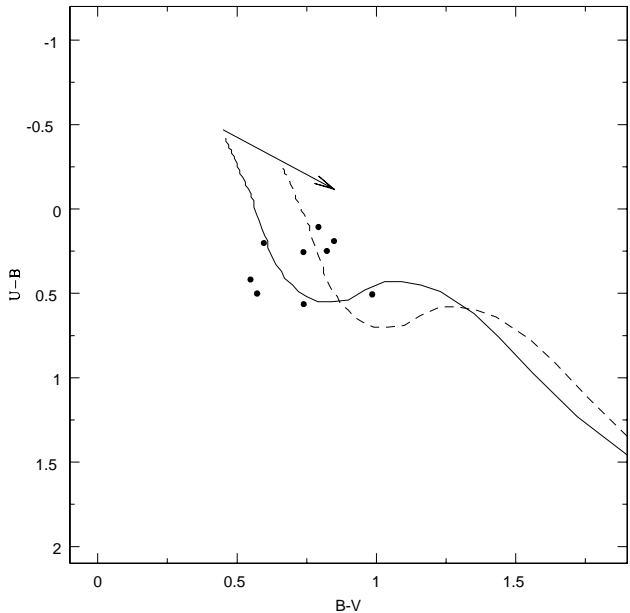}
\caption{$(U-B)/(B-V)$ TCD for variables associated with the BP population. The $UBV$ photometric data are taken from
Jose et al. (2008). The continuous and dashed curve show the ZAMS by Girardi et al. (2002) which are moved in the direction of the reddening vector for reddening $E(B-V)= 0.70$ mag and 0.90 mag. The straight line indicates reddening vector for $E(U-B)/E(B-V)$ = 0.72.} 
\end{figure}

\begin{figure}
\includegraphics[width=7cm]{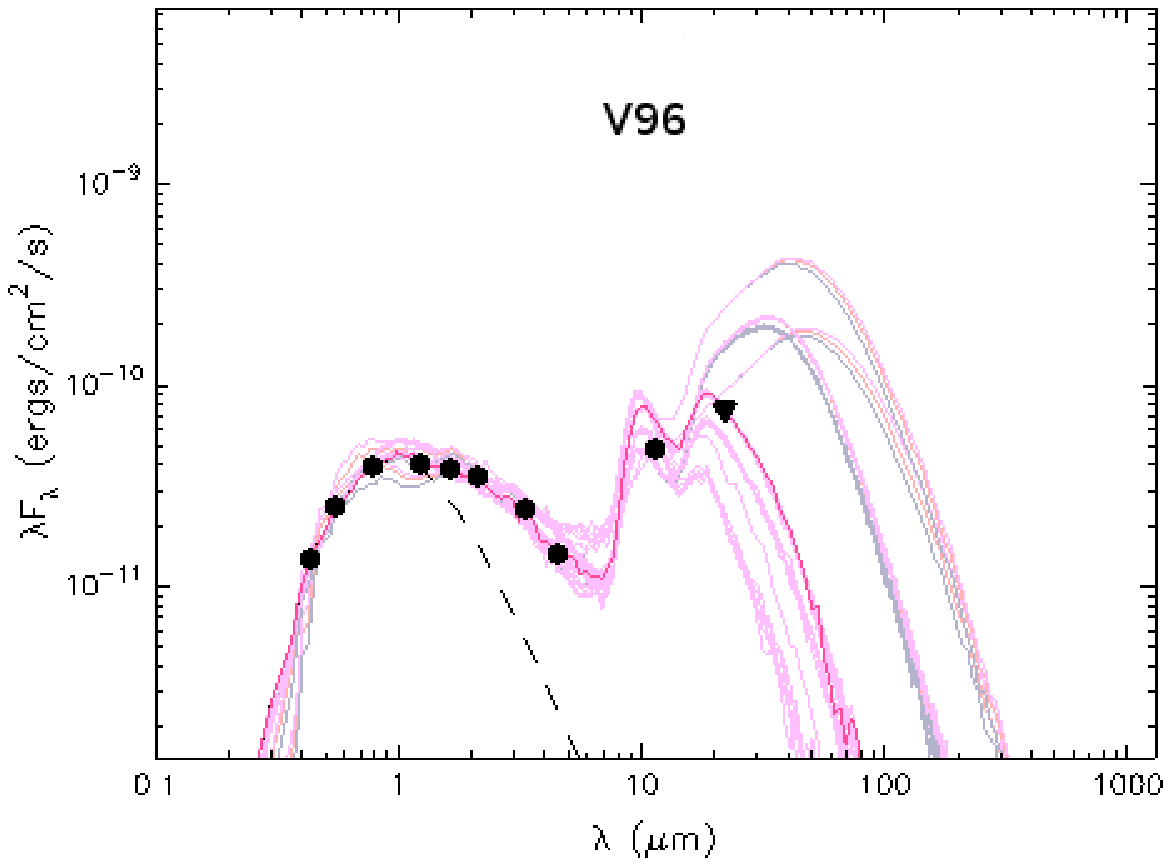}
\includegraphics[width=7cm]{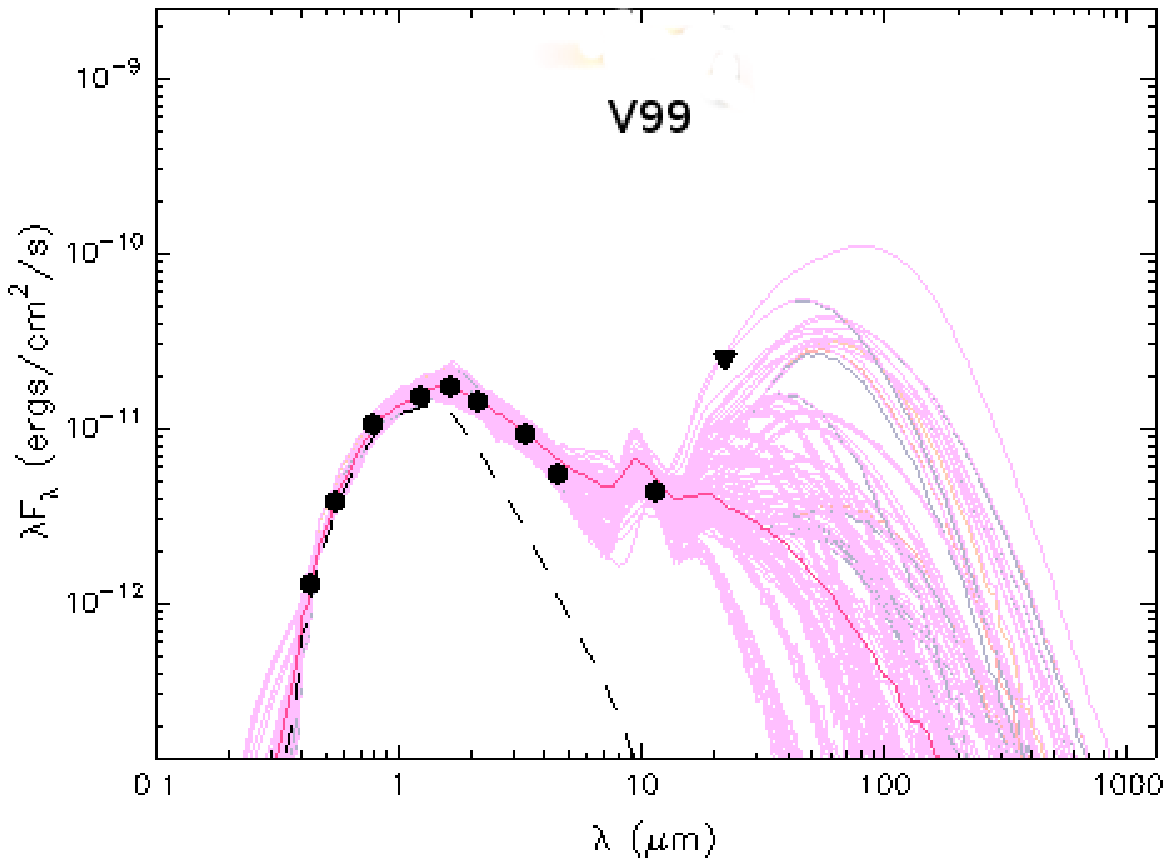}
\caption{SEDs of the YSOs V96 and V99.
The dark line shows the best fit model, and the light pink lines show subsequent models
that satisfy $\chi^2 - \chi^2_{\rm min} \leq 2N_{\rm data}$ criterion. The
dashed line represents the best fit model for the stellar photosphere of
central source. The observed flux values are represented by filled circles. 
 The 22 $\micron$ flux is shown by triangle, which is considered as upper limit while fitting the model.}
\end{figure}

\subsection{Transformation, identification of variable and determination of period}
The corrected $v$ instrumental magnitudes of variables were converted into
standard ones by obtaining transformation equation using the photometric data for the region by Jose et al. (2008).
The following transformation equation was used to convert present $v$ magnitudes into standards ones.
\begin{eqnarray}
V = (0.965\pm0.01)\times v+1.453\pm0.022   \nonumber\\
\end{eqnarray}
where $V$ and $v$ are standard and 
instrumental magnitude, respectively.

The $g^{\prime}$ band instrumental magnitude was converted into $V$  band standard magnitude using the following
transformation equation
\begin{eqnarray}
V = (0.912\pm0.021)\times g_{inst}^{\prime}+0.79\pm0.032   \nonumber\\
\end{eqnarray}

We could  not use colour term  in the transformation equations as time series photometry for two colours was not available. However, to check the effect of colour term on transformation, we used average $v$ magnitude from present time series observations and converted to standard magnitude using the following equation 
\begin{eqnarray}
V-v=(-0.079\pm0.012)\times(V-I)+(1.085\pm0.017)  \nonumber\\
\end{eqnarray}
where $V$ and $(V-I)$ colours are taken from Jose et al. (2008). We have found that the effect of colour term in one magnitude bin is negligible.
Since in the present study the amplitude is less than one magnitude, it will
have no effect on light curves of variable stars.

The light curves of all the cross matched stars by the DAOMASTER  were generated by plotting standard $V$ magnitudes of stars against Julian date (JD). 
We visually checked the light curves of all the 1721 stars.
A star has been selected as variable star if it showed brightness variation of at least $\sim$0.01 mag. The present study did not use RMS dispersion criterion as this 
criterion may not able to detect small amplitude periodic variables. Hence, the probable variables were identified visually by inspecting their light curves.
The visual inspection yields 130 variable candidates in the field of Stock 8. 
The sample light curves of a few variables are shown in Fig. 3.
The sample of data for variables is given in Table 2. The complete table is available in electronic form. 
The variable candidates identified in the present work are marked in Fig. 1.
The variability of variable stars was also checked using the $Chi$ square test (Sesar et al. 2007). Out of 130 variables discovered, 112 have probability $\ge$ 90\%. Visual inspection of the remaining variables reveals a significant variability, hence we also considered them suspected variables.
The identification number, coordinates and optical as well as photometric data in NIR of the identified variables are listed in Table 3. To convert CCD pixel coordinates of the identified variables to their celestial coordinates (RA and DEC) for J2000 we have used the CCMAP and CCTRAN tasks available in IRAF.

The Lomb-Scargle periodogram (Lomb 1976; Scargle 1982) has been used to determine the probable periods of variable stars. This periodogram gives better results even when data are taken at
irregular intervals. The periods were further confirmed using the NASA exoplanet archive periodogram service. The phased light curves were visually inspected, and we 
have considered the period which shows the best phased light curve. We have listed the most probable periods of the stars with their amplitudes in Table 3. The folding of light curves for variable stars are done with their estimated periods. 
The phased light curves will be further discussed in Section 6.

\section{Identification of probable members of Stock 8}
The $V/(V-I)$ CMD of the cluster region by Jose et al. (2008, cf. their figure 8b and 8c) clearly reveals contamination due to field star population. 
Ninety and 39 of the 130 variables identified in the present work are found to be common with those reported by Jose et al. (2008) and (2017), respectively. All the 39 variables (Jose et al. 2017) are included in the data by Jose et al. (2008).  
The individual membership of these 90 stars was not discussed by Jose et al. (2008). Their figure 8d represents statistically cleaned CMD of the cluster. Hence, we have used $(U-B)/(B-V)$, NIR ($J-H)/(H-K)$ TCDs and $V/(V-I)$ CMD to find out the association of the identified variables with the cluster. 

The $(U-B)/(B-V)$ TCD for 44 variable candidates is shown in Fig. 4 as
$(U-B)$ colours are available for only 44 variables. 
The distribution of MS variables in Fig. 4 reveals a variable
reddening in the cluster region with the reddening value of $E(B-V)_{min}=0.40 $ mag.
The sources lying within the MS band having $E(B-V)$ = 0.4 mag to 0.6 mag with spectral type from O to A can be considered 
as possible MS members of the cluster.
 Fig. 5 shows $(J-H)/(H-K)$ TCD of 101 variables as the $JHK$ counterparts
for 29 variables could not be detected. 
The YSOs either show significant amount of H$\alpha$ emission and/or NIR excess. Therefore, the $(J-H)/(H-K)$ TCD is a very useful tool to identify YSOs. 
In Fig. 5, the sources lying in `F' region could be either field stars or  Class III and Class II sources with small NIR excesses. The sources lying in the `T' region can be considered mostly as CTTSs (Class II objects). 

Fig. 6 shows the $V/(V-I)$ CMD for 90 variables as $(V-I)$ colour
for remaining 40 variables was not available.
We have plotted a theoretical isochrone of 4 Myr for Z=0.02 (Girardi et al. 2002) with continuous curve in Fig 6. Fig. 6 also plots the PMS isochrones and evolutionary tracks for various ages and various masses (Siess et al. 2000), respectively. The distance (2.05 kpc) and minimum reddening $E(V-I)_{min}=0.5$ mag have been used to correct all the isochrones and evolutionary tracks. The $E(V-I)_{min}$ has been estimated using the relations $E(V-I)/E(B-V )$ = 1.25 mag and $E(B-V )$ = 0.40 mag.

Based on the above mentioned TCDs and CMD, we have established membership of 51 stars (28 MS and 23 PMS stars). 
Of the remaining 79 stars, 29 variables remained unclassified due to unavailability of their photometric and NIR data. 
Fifty stars may belong to the field star population. Twenty one out of these 50 stars are found to be distributed below the 10 Myr isochrone around $(V-I)$ $\sim$ 1.2 mag and $V$ $\sim$ 15.6 to 17.0 mag. This population may belong to
the blue plume (BP) of Norma-Cygnus arm as discussed by Jose et al. (2008).
$U-B$ colour is available for 9 of these 50 variables. The location of these 9 variables in $U-B/B-V$ diagram (Fig. 7) indicates 
that these stars have $E(B-V)$ in the range of 0.7-0.9 mag, comparable to the BP Population of Norma-Cygnus arm (see e.g. Pandey et al. 2006 
and Jose et al. 2008). Hence, we consider these 9 variable stars as of the BP population. 
The classification of variables detected in the present work is given in the
last column of Table 3.  

\section {Age and mass estimation}
The ages and masses of PMS stars have been estimated by comparing present observations with the theoretical models to study the evolution of amplitude and period of PMS stars. For this, we have used $V/V-I$ CMD and PMS isochrones by Siess
et al. (2000). The PMS isochrones of Siess et al. (2000) in the age range of 0.1 to 10 Myr with an interval of 0.1 Myr have been used.  We then interpolated these isochrones to make more continuous curves. These theoretical isochrones of Siess et al. (2000) corrected for the distance (2.05 kpc) and reddening ($E(V-I)_{min}=0.5$) have been compared with the location of PMS stars in the $V/V-I$ CMD. Finally, we determine the age and mass of the PMS star corresponding to the closest isochrone on the CMD. Thus, the estimated age and mass   
may be affected by random errors in observations, errors in transformation to the standard system and systematic errors because of using different theoretical evolutionary tracks. We presume that the systematic errors do not have any effect on the mass and age estimates obtained in the present work as we are using the model by Siess et al. (2000) for all the PMS stars. Assuming normal error distribution and using the Monte Carlo simulations (see e.g., Chauhan et al. 2009) the random errors were propagated to the observed estimates of $V/(V-I)$ and $E(V-I)$ to estimate the random errors in the determination of mass and age. Another source of error may be the presence of binaries. A star will become brighter in the presence of binary which, consequently, will yield a younger age. For equal mass binary, the expected error in the estimation of age of PMS stars is $\sim$50 to 60\%. Since we do not know the fraction of binaries in the Stock 8 cluster, it is difficult to determine the effect of binaries on the estimation of mean age. However, the study of 
Duquennoy \& Mayor (1991) of multiplicity among solar type stars in the solar neighbourhood suggests that the distribution of mass ratio (M1/M2) shows a peak around $\sim$ 0.23. If the binaries in Stock 8 cluster have a similar mass distribution, the effect of binaries on age estimates may not be significant. 
The age and mass estimates of the YSOs are in the range of 0.2 to 5.8 Myr and  $\sim$0.5 to $\sim$2.75 $M_{\odot}$, respectively,  which are comparable with the ages and masses of TTSs. The age spread of YSOs
in the Stock 8 region indicates a non-coeval star formation in the region.
The estimated ages and masses along with their errors are given in Table 4. 

\section {Spectral Energy Distribution}

The spectral energy distribution (SED) of YSOs is a useful tool to characterize the circumstellar disk properties of YSOs. To construct the SEDs of YSOs we have used Robitaille et al. (2006, 2007) models for radiative transfer and multiwavelength i.e., optical ($BVI$), NIR ($JHK$) and WISE (3.4, 4.6, 12.0, and 22.0 $\micron$) data. The 22 $\micron$ data have been used as upper limit because of its large beam ($\sim$22 arcsec) and crowding of the region. To fit the SED models we have taken distance to the cluster as 2.05 $\pm$ 0.10 kpc and also used a maximum value of extinction ($A_{V}$) which was determined by tracing back the current location of the YSOs on $J/(J-H)$ diagram to the intrinsic locus of dwarf along the reddening vector (Samal et al. 2010). The minimum value of foreground extinction in the direction of Stock 8 is taken as 1.2 mag.

We have fitted SEDs for 11 identified YSOs. The WISE data for stars with IDs V19, V26, V37, V45, V84, V85, V90, V91, V110, V113, and V128 were not available. The SEDs of two sources V96 and V99 are shown in Fig. 8 as an example. As expected, the SED models show a high degree of degeneracy in the absence of mid/far
infrared  and millimeter data, however, the SEDs of 4 identified CTTSs (V106, V99, V125 and V61) and 2 probable HAeBe sources indicate the presence
of NIR/MIR-excess emission, possibly due to circumstellar disk. It is not possible to characterize all the SED parameters from the model due to
limited data points. However, the SED models fit the observed data fairly well in the wavelength range from 0.4 $\micron$ to 12 $\micron$, and  
we expect that the age and mass estimations of the YSOs should be constrained well enough. Table 4 lists the age and mass estimated from the SED analysis.
These parameters have been obtained using the criterion $\chi^{2}$-$\chi^{2}_{min}$$\le$ 2N$_{data}$ weighted by
e$^{({{-\chi}^2}/2)}$ of each model as done in Samal et al. (2012), where  $\chi^{2}_{min}$ is the goodness-of-fit parameter for the best-fit model and $N_{\rm data}$
is the number of input observational data points. Table 4 indicates that age and mass estimations using the SED models are higher by $\sim$2 times
in comparison to the estimates based on the $V/(V-I)$ CMD.
Since photometry in optical bands in comparison to NIR bands has better accuracy and variables studied in the present study have relatively lower extinction, the age and mass estimated from $V/(V-I)$ CMD seem to be reliable. 
Table 4 also indicates that 
the mass and age estimates using $V/(V-I)$ CMD have lesser uncertainties in comparison to those obtained 
from the SED fitting.

\begin{table}
\caption{The effective temperature ($T_{\rm eff}$), bolometric correction ($BC$),  bolometric magnitude ($M_{bol}$), luminosity (L).}
\scriptsize
\begin{tabular}{llccl}
\hline
ID&$\log T_{\rm eff}$&   $BC$ &  $M_{bol}$ & $\log(L/L_{\odot})$\\
  &              &   (mag) & (mag)   &                     \\
\hline
         V23&  3.996 &  -0.07031 &  0.02169 & 1.884 \\
         V38&  4.060 &    -0.4844&  -0.6554 & 2.155 \\ 
         V42&  4.165 &    -1.125 &    -1.26 & 2.397 \\
         V55&  4.025 &   -0.2422 &   0.8918 & 1.536 \\
         V56&  4.235 &    -1.508 &    -2.42 & 2.861 \\
         V59&  4.123 &   -0.8281 &   0.4519 & 1.712 \\
         V64&  4.054 &   -0.3359 &   0.9061 &  1.53 \\
         V66&  3.963 &   0.01562 &    1.394 & 1.335 \\
         V68&  4.073 &   -0.6094 &    -1.18 & 2.365 \\
         V76&   3.95 &    0.1406 &   0.1466 & 1.834 \\
         V79&  4.029 &   -0.2266 &   0.6904 & 1.617 \\
         V80&  3.987 &   -0.0625 &   0.3485 & 1.753 \\
         V81&  3.997 &  -0.04688 &     1.55 & 1.273 \\
         V93&  3.977 &  -0.01562 &  -0.4896 & 2.089 \\
         V94&  4.234 &    -1.438 &   -2.048 & 2.712 \\
        V103&  4.056 &   -0.5156 &  -0.1916 & 1.969 \\
        V105&  4.093 &   -0.5938 &  0.03275 & 1.906 \\    
        V107&  4.057 &   -0.3984 &  0.09256 & 1.856 \\
        V109&  4.294 &    -1.727 &     -2.1 & 2.733 \\
        V114&  4.071 &      -0.5 &    0.329 & 1.761 \\
        V116&  4.104 &   -0.7578 &  -0.2288 & 1.984 \\
        V120&  4.068 &   -0.5391 &  -0.1251 & 1.943 \\
        V121&  4.076 &   -0.5391 &   0.9369 & 1.518 \\
        V130&  4.411 &     -2.25 &   -6.097 & 4.332 \\        
\hline
\end {tabular}
\end{table}

\begin{figure}
\includegraphics[width=9cm, height=9cm]{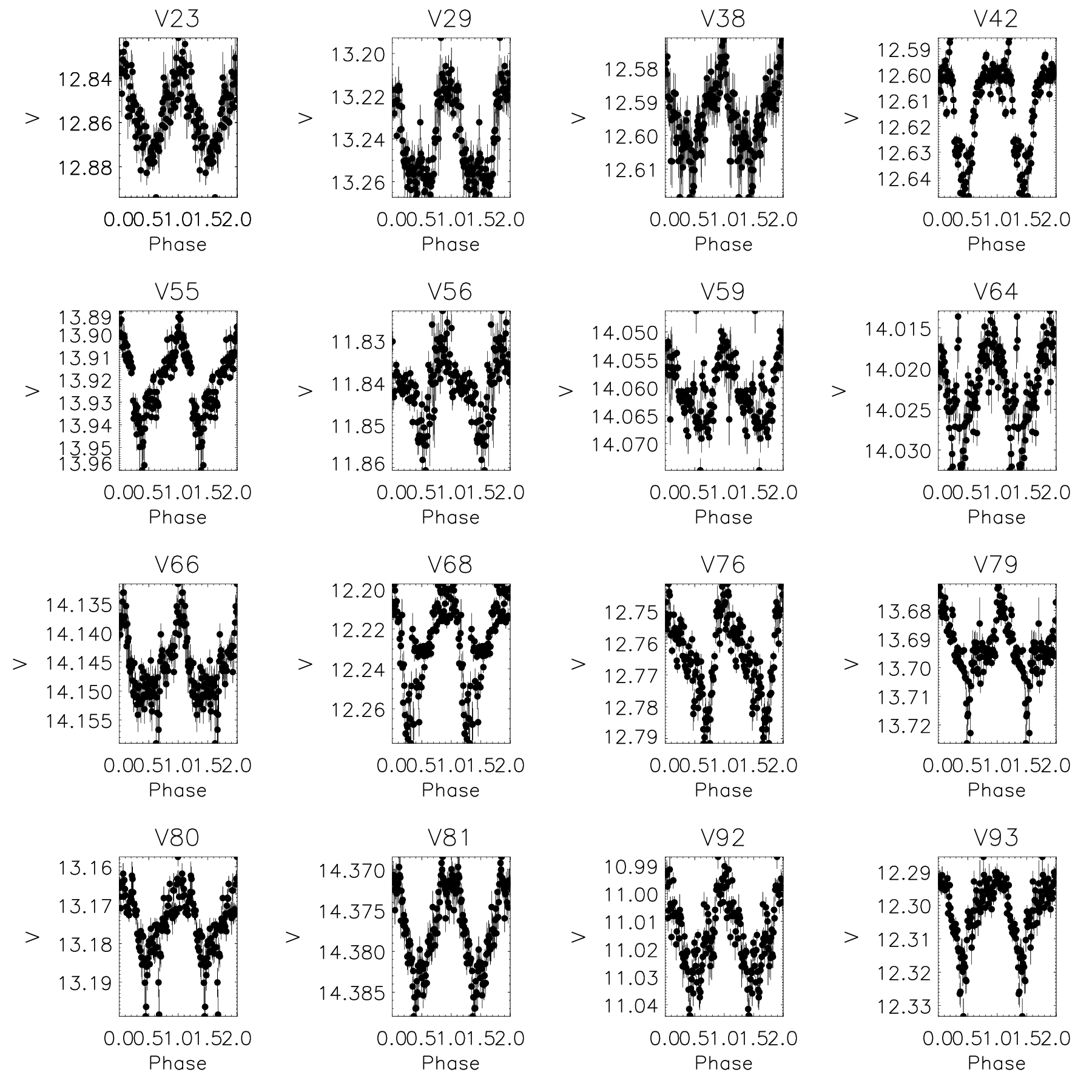}
\includegraphics[width=9cm, height=9cm]{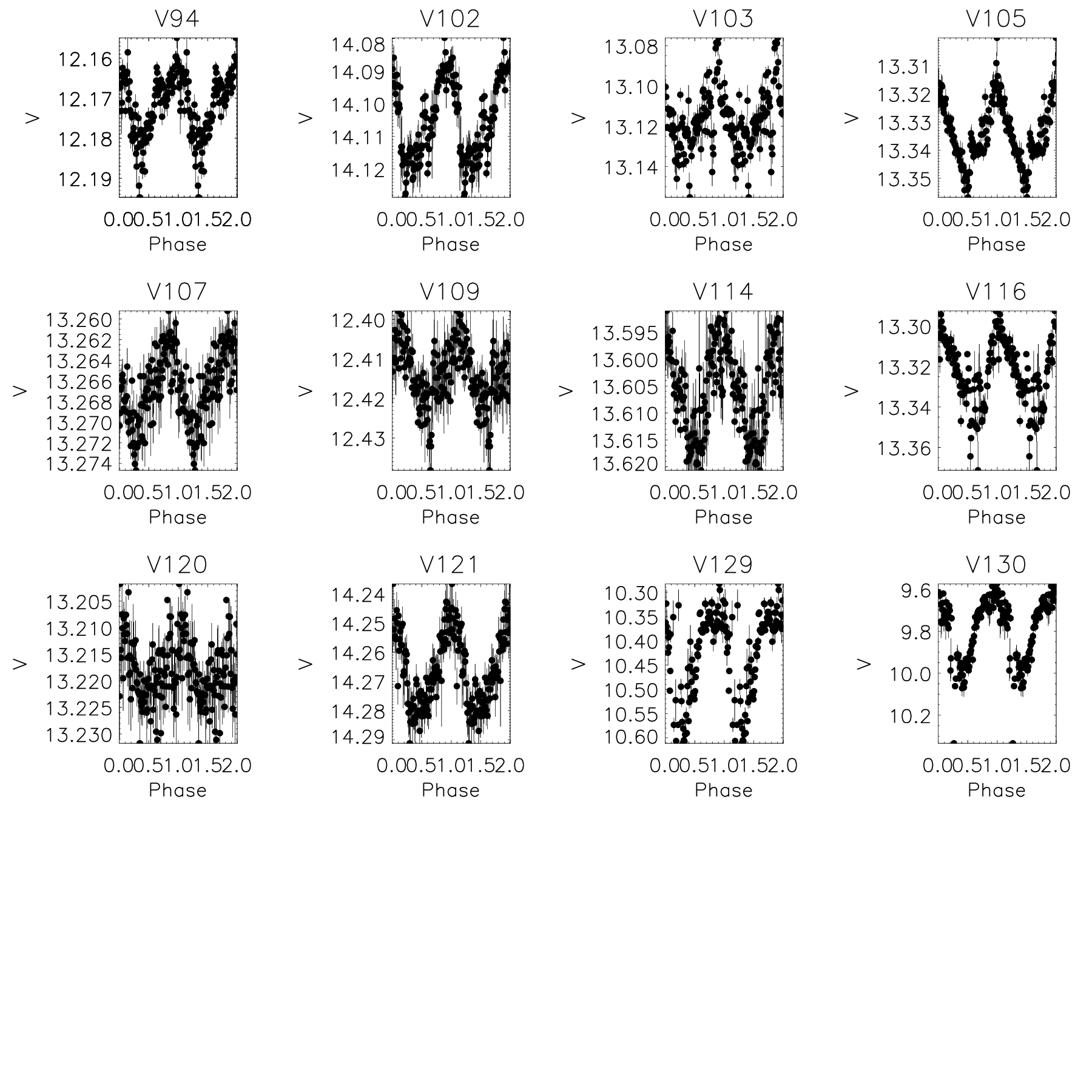}
\caption{The $V$ band phased light curves of MS variable stars. }
\end{figure}

\begin{figure}
\includegraphics[width=9cm, height=9cm]{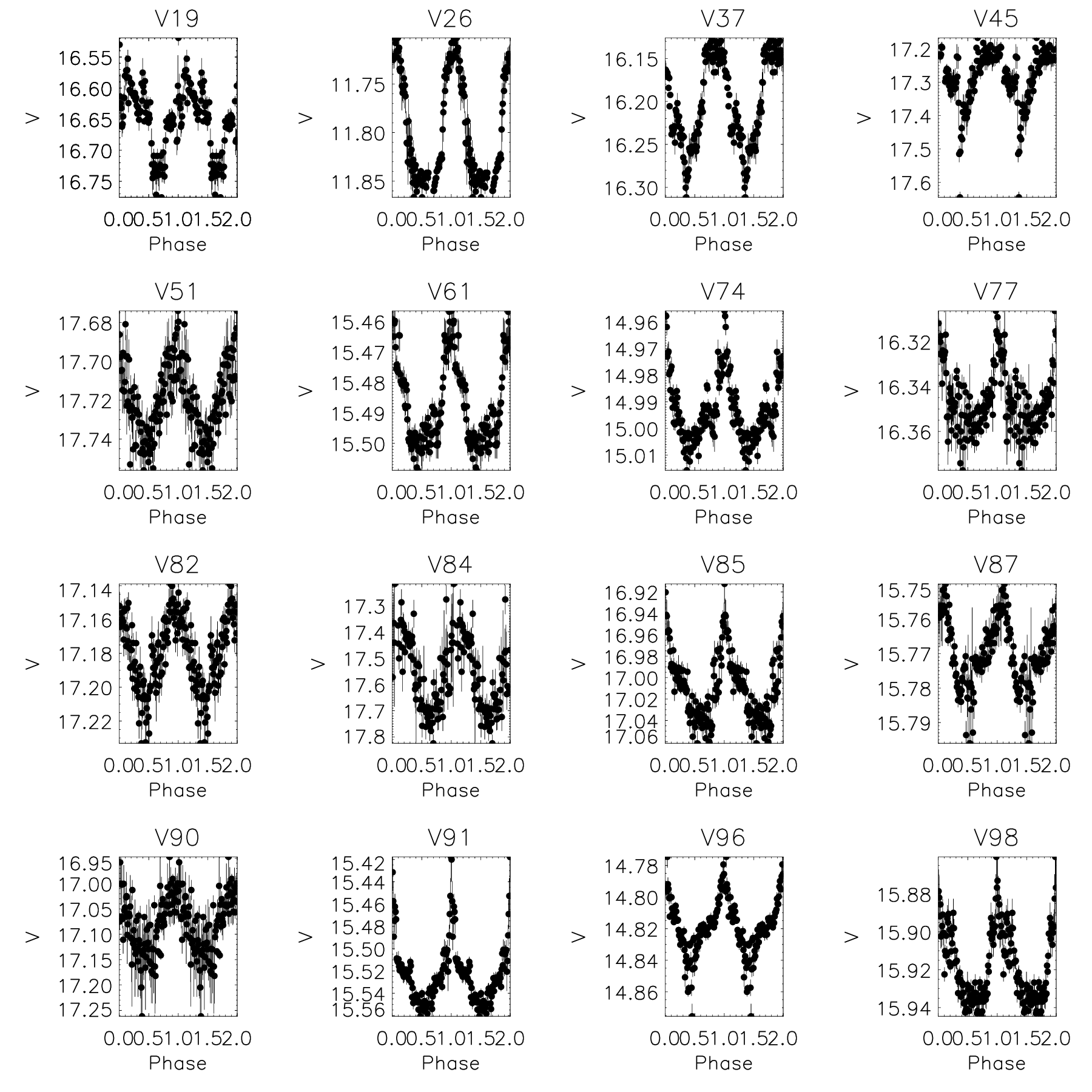}
\includegraphics[width=9cm, height=9cm]{phase_pms2.ps}
\caption{The $V$ band phased light curves of PMS variable stars.} 
\end{figure}

\begin{figure}
\includegraphics[width=9cm, height=9cm]{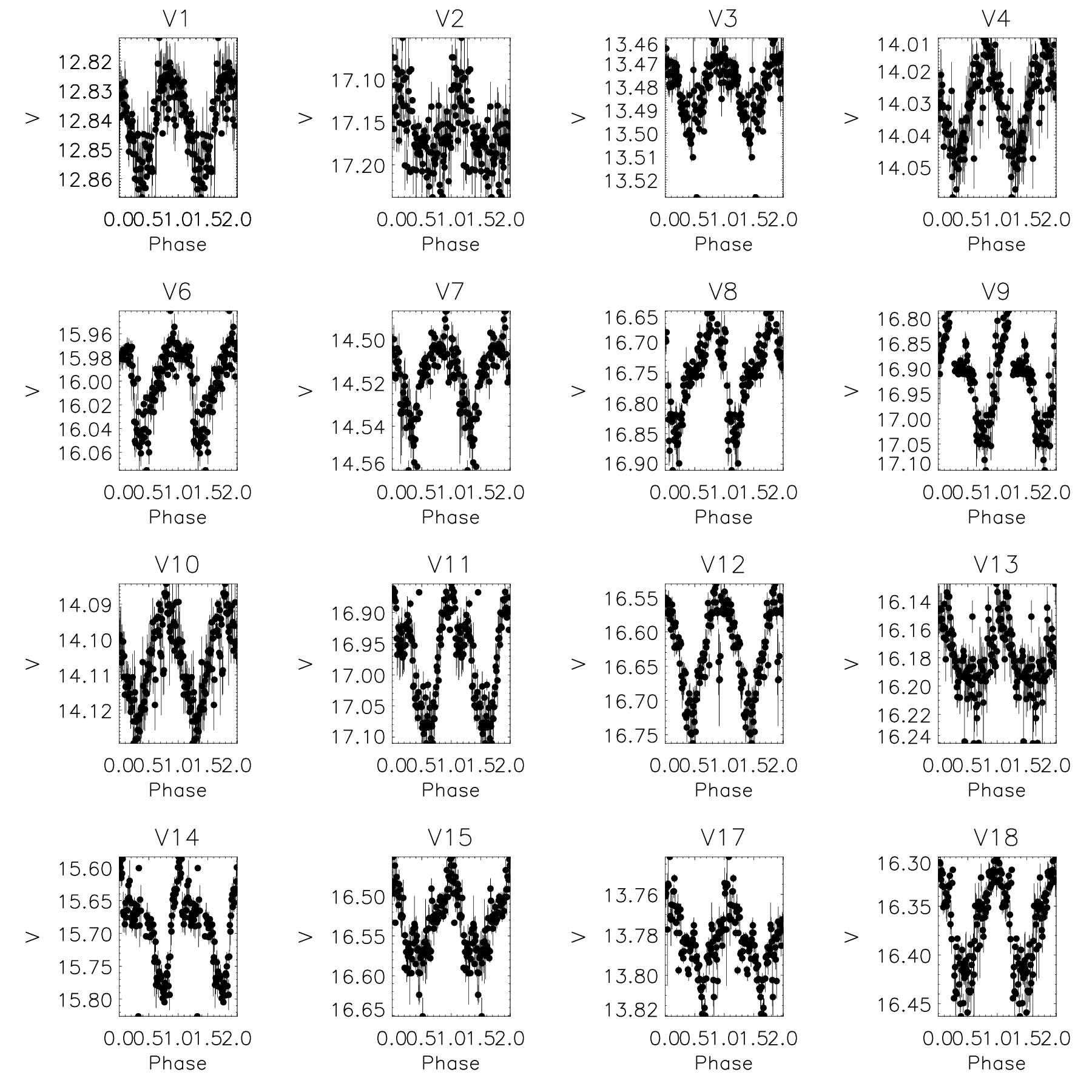}
\includegraphics[width=9cm, height=9cm]{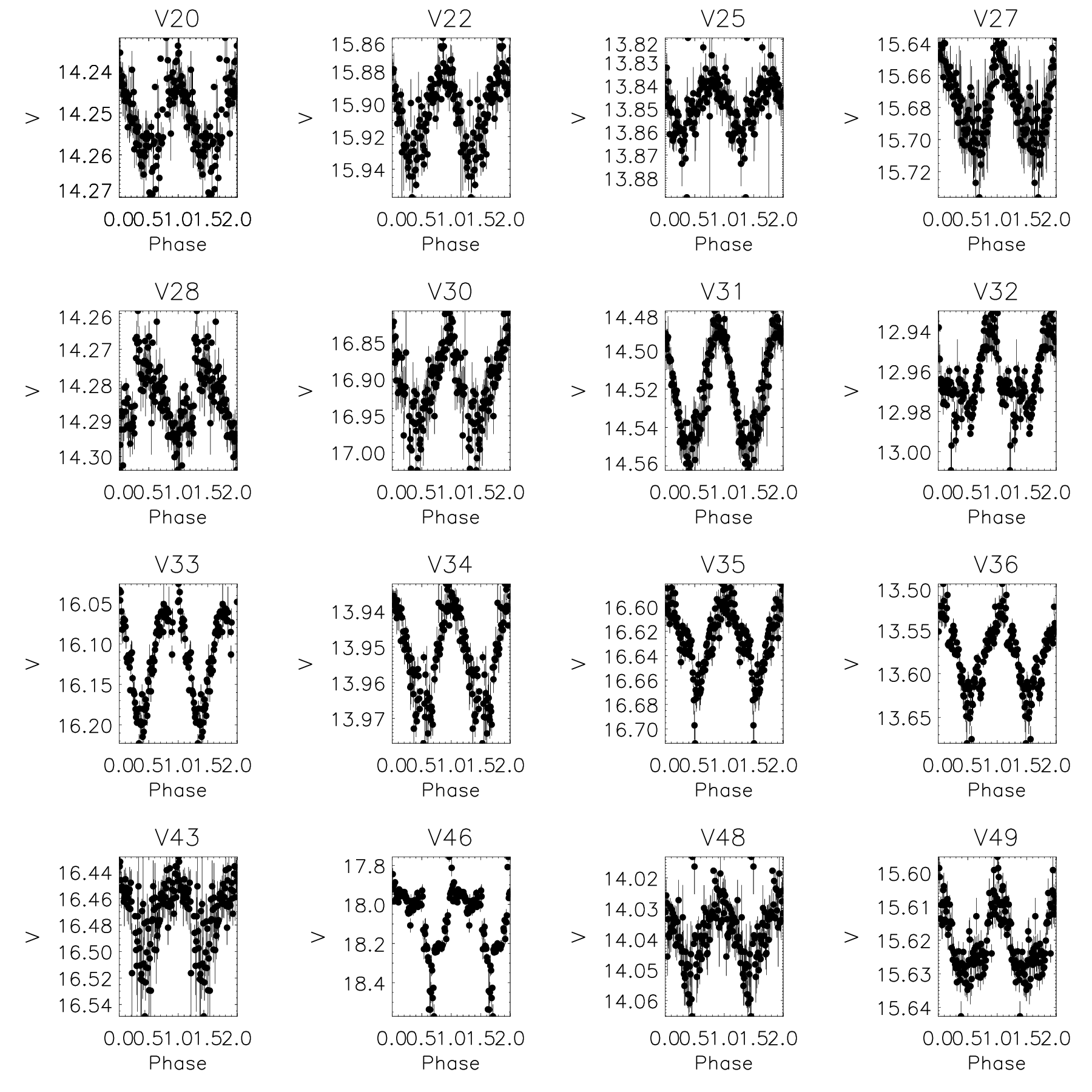}
\caption{The $V$ band phased light curves of unclassified and field variables.}
\end{figure}

\setcounter{figure}{10}
\begin{figure}
\includegraphics[width=9cm, height=9cm]{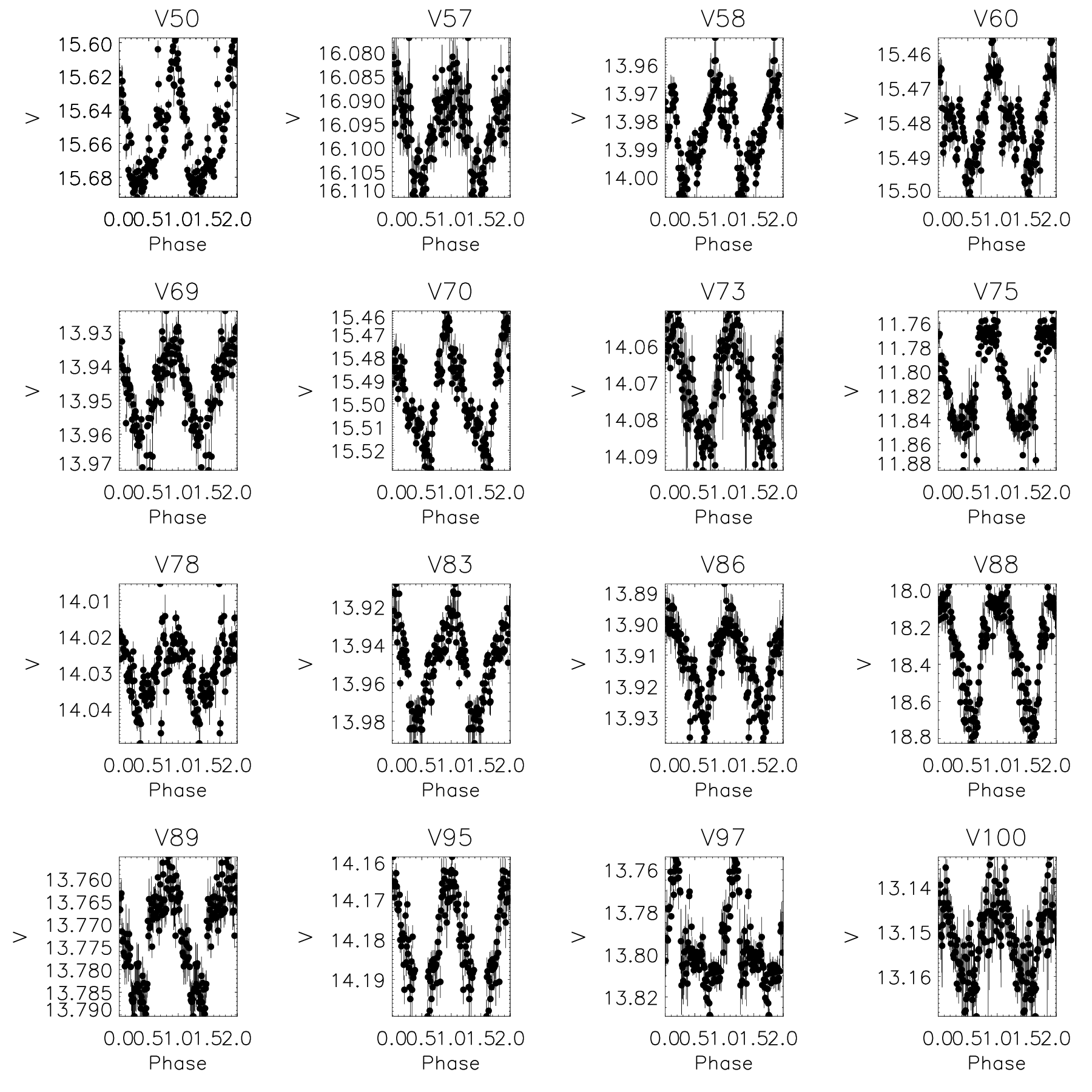}
\includegraphics[width=9cm, height=9cm]{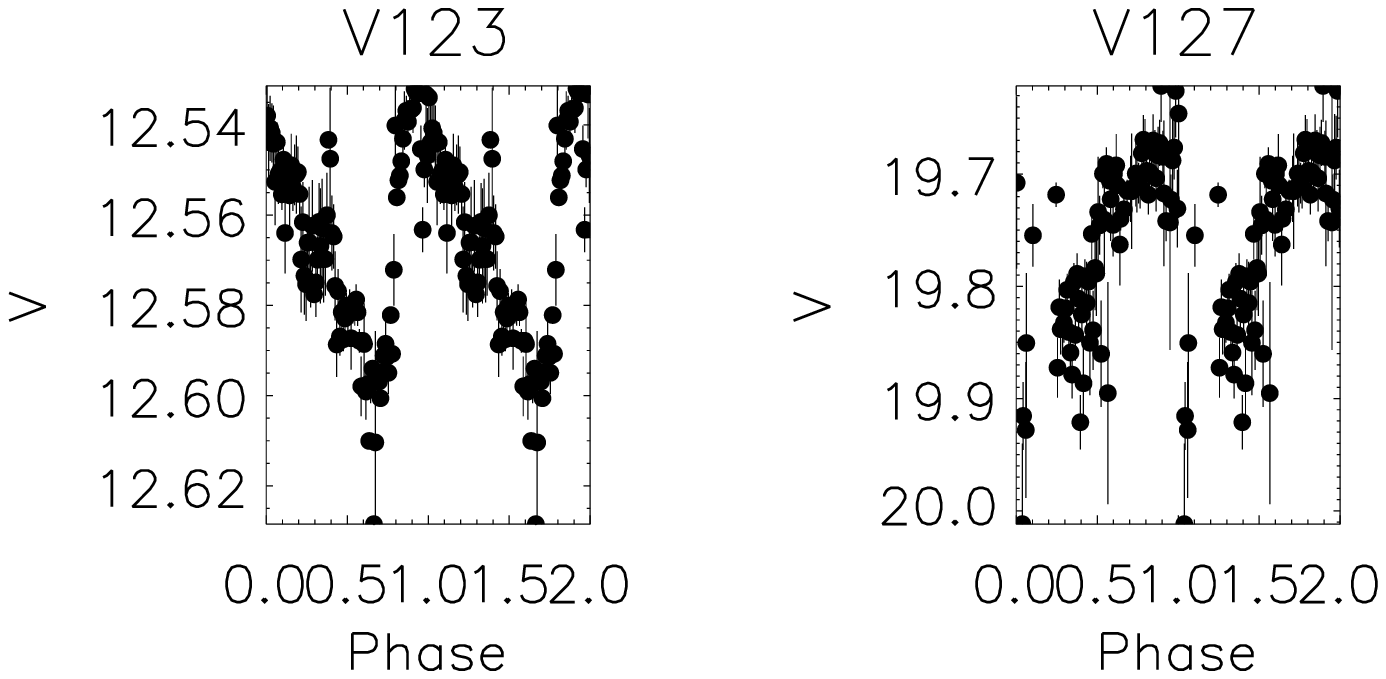}
\caption{continued.}
\end{figure}

\begin{figure}
\includegraphics[width=9cm, height=9cm]{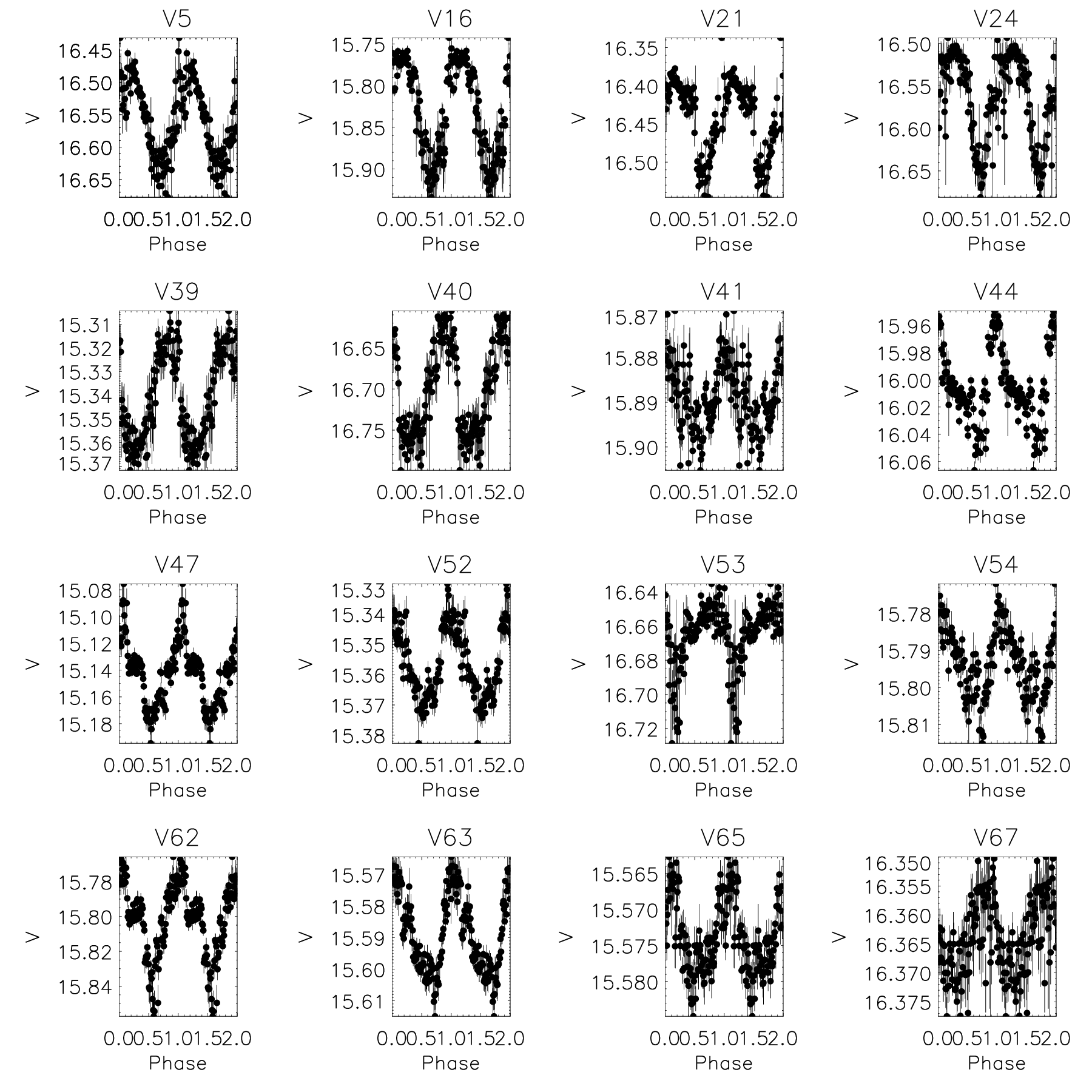}
\includegraphics[width=9cm, height=9cm]{phase_bp2.ps}
\caption{The $V$ band phased light curves of variable stars associated to the BP population.} 
\end{figure}

\section{Nature of variable stars}
Figs. 9, 10, 11 and 12 display phased light curves of variable stars identified as the MS stars, PMS population of cluster, field population and unclassified, and
probable BP population of Norma-Cygnus arm, respectively, where the average magnitude in 0.01 phase bin is represented by filled circles and error bars represent standard deviation of mean magnitude in particular bin.

\begin{figure}
\includegraphics[width=9cm]{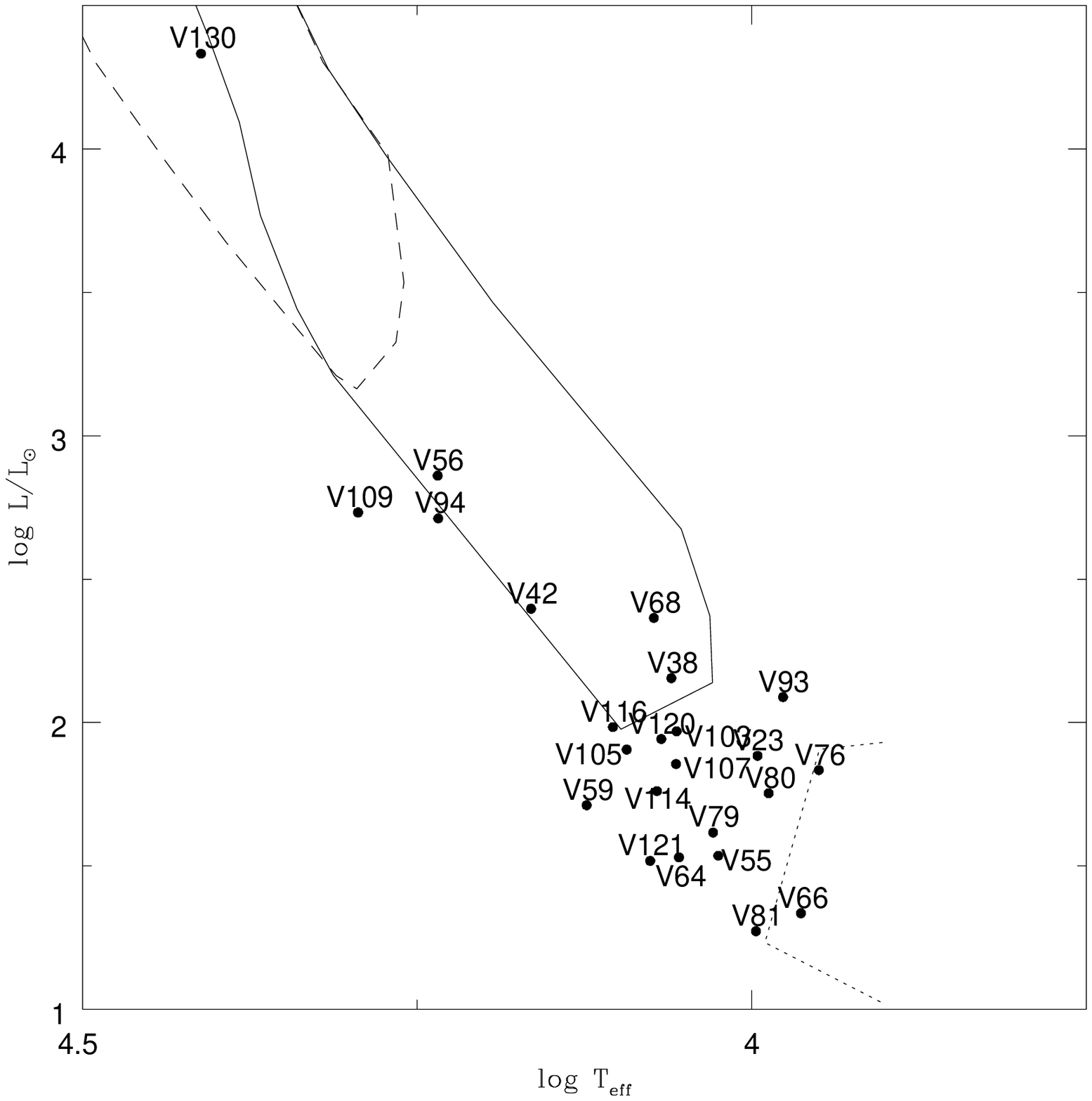}
\caption{ $\log(L/L_{\odot})/ \log T_{\rm eff}$ diagram of the cluster Stock 8 for the probable MS variable stars. 
The continuous and dotted curve are the instability strip for SPB and  $\delta$ Scuti stars, respectively and dashed curve shows the location of $\beta$ Cep stars (for references see Balona et al. 2011).
}
\end{figure}
\subsection {MS Variables}
In order to characterize the MS variables associated with the cluster region, we have plotted them in H-R diagram (Fig. 13). 
Fig. 13 shows the log($L/L_{\odot}$ )/ log$ T_{\rm eff}$ diagram (H-R diagram) for the MS variables. In Fig. 13, we could not plot 4 MS variables V29, V92, V102 and V129 due to unavailability of their $U-B$ colours. The luminosity and temperature of a star is determined as follows. First of all, we have determined the intrinsic $(B-V )$ colours 
 with the help of the $Q$-method (Gutierrez-Moreno 1975).
The absolute magnitude, $M_{V}$ was obtained assuming distance modulus of the Stock 8 as 12.8 mag.
To convert $M_{V}$ into luminosity ($\log L/L_{\odot}$) it is necessary to know the  bolometric correction ($BC$).
The $BC$ is determined from the effective temperature as it is a function of
temperature. 
The effective temperature $T_{\rm eff}$ was determined using the relation between $T_{\rm eff}$ and intrinsic $(B-V)$ colour by Torres (2010). 
The $BC$ has been calculated using $T_{\rm eff}$ with the help of Torres (2010) relation.  After that luminosity of the stars was determined from a relation $\log(L/L_{\odot} ) = −0.4(M_{bol}-M_{bol_{\odot}})$, where $M_{bol} = M_{V} + BC$, and $M_{bol}$ and $M_{bol_{\odot}}$ are bolometric magnitudes for the star and Sun, respectively. In the case of Sun, the bolometric magnitude $M_{bol_{\odot}}$ was considered as 4.73 mag (Torres 2010). 
Thus, the determined parameters such as  $T_{\rm eff}$, $BC$, $M_{bol}$ and $\log L/L_{\odot}$ for MS stars are given in Table 5.

Fig. 13  also shows the theoretical slowly pulsating B (SPB) instability strip (continuous curve), location of $\beta$ Cep stars (dashed curve) and empirical $\delta$ Scuti instability strip (dotted curve) taken from Balona et al. (2011; references therein). Twenty eight variables are found to be MS type stars associated with the cluster. The periods of these stars estimated in the present work range between 0.053 d to 0.450 d and amplitudes range from 0.01 to 0.38 mag.

 In the H-R diagram (cf. Fig. 13), the location of MS variable stars indicates that 7 stars (V38, V42, V56, V68, V94, V109 and V116) could be SPB stars, whereas star V130 could be a $\beta$ Cep variable. Two stars V66 and V76 are found to lie in the $\delta$ Scuti region and these could be $\delta$ Scuti variables. Fourteen stars V23, V55, V59, V64, V79, V80, V81, V93, V103, V105, V107, V114, V120, and V121  are located in the gap present between SPB and $\delta$ Scuti instability region. In the case of the open cluster NGC 3766, Mowlavi et al. (2013) have found a large population of new variable stars between SPB stars and the $\delta$ Scuti stars, the region where no pulsations were expected on the basis of theoretical models. The findings of Mowlavi et al. (2013) were further supported by Lata et al. (2014) in the case of young cluster NGC 1893. Mowlavi et al. (2013) have reported periods of these variable stars in the range of 0.1 to 0.7 d, with amplitudes between 1.0  to 4.0 mmag, whereas Lata et al. (2014) found the periods of a new class of variables ranging from 0.17 to 0.58 d with an amplitude of variation in the range 0.007 to 0.019 mag. Periods and amplitudes of these new class variables identified in the present work have range of 0.064 to 0.364 d and $\sim$0.01 to $\sim$0.04 mag, respectively. 

The origin of variability of these stars could be pulsation. One of the probable causes of pulsation in these stars could be rapid rotation which alters the internal conditions of a star enough to sustain stellar pulsations. Another cause for the brightness variation in these stars might be the presence of spots on the surface of such rotating stars and that these spots would induce light variations as the star rotates. But hot stars are not expected to be active, and no theory can currently explains how spots could be produced on the surface of such stars. Balona et al. (2011) analyzed light curves of 48 B-type stars observed by Kepler and did not find any star lying between the red end of the SPB stars and the blue end of $\delta$ Scuti type stars.

\subsection {PMS variables}
In the present work, we classify 23 stars as probable PMS stars (cf. Table 3). The identified PMS variables have ages and masses in the range of 0.2 - 5.8 Myr and 0.5 - 2.75 $M_{\odot}$ respectively, which are comparable to those of TTSs. The majority of these stars are located in `F' region of $(J-H)/(H-K)$ TCD. As discussed in Section 3, the `F' region may contain some Class II sources which have small NIR excess. In fact, figure 3a by Pandey et al. (2014) in the case of the young open cluster NGC 1893 reveals that the `F' region contains a significant amount of Class II sources. A careful view of figure 3b by Pandey et al. (2014) manifests that almost all the Class III YSOs are located toward the left side of the `F' region, whereas a significant number of Class II YSOs can be noticed towards the right side of the `F' region. Hence, we subdivided `F' region, shown with a dot-short dash line in Fig. 5, into `F1' and `F2' regions. The PMS stars lying in the `F2' region are also considered as Class II sources.

The star V125 lies in the `T' region in $(J-H)/(H-K)$ TCD (Fig. 5) hence should be a CTTS, likewise stars V61, V99, V106 and V110 that lie in the `F2' region. The amplitudes of stars V61, V99, V106 and V110 are in the range from 0.026 - 0.412 mag, whereas amplitude of star V125 is found as 0.360 mag. The periods of these CTTSs are in the range of 0.128 - 0.648 d. The location of stars V96 and V98 suggests that these could be HAeBe stars. The masses and ages of these sources are 2.13 $M_{\odot}$, 2.75 $M_{\odot}$  and 3.13 Myr, 1.92 Myr, respectively. The period and amplitude of V96 and V98 are estimated as 0.159 d, 0.121 d and 0.051 mag, 0.042 mag, respectively. In the case of NGC 1893, Lata et al. (2012) have identified 2 probable HAeBe stars with periods 0.30 d and 0.47 d and amplitudes 0.01 mag and 0.02 mag. Stars numbered V19, V26, V37, V45, V51, V74, V77, V82, V84, V85, V87, V90, V91, V113, V126, and V128 lie in the `F1' region and could be WTTSs. These WTTSs have amplitude in the range of 0.013 - 0.456 mag and periods lie in the range of 0.095 - 0.498 d. 

The present sample of TTSs manifests that CTTSs have amplitude in the range of  0.039 - 0.288 mag, whereas the amplitude of majority of  WTTSs varies from 0.013 to 0.242 mag. This indicates that the brightness of CTTSs varies with larger amplitude in comparison to WTTSs. This result is in agreement with the previous studies (Grankin et al. 2007, 2008; Lata et al. 2011, 2012 and 2016).
As discussed, e.g., by Carpenter, Hillenbrand \& Skrutskie (2001) the larger amplitude in the case of CTTSs could be due to the presence of hot spots on the stellar surface produced by an accretion mechanism. Hot spots cover a small fraction of the stellar surface but with a higher temperature causing larger amplitude of brightness variations. The smaller amplitude in WTTSs suggests dissipation of their circumstellar discs or these stars might have cool spots on their surface which are produced due to convection and differential rotation of star and magnetic field. The results obtained in the present study are in agreement with work reported by Grankin et al. (2007, 2008). These results also match with that of Lata et al. (2011, 2012 and 2016).

\subsection {Field, Blue Plume Population in Norma-Cygnus arm and non classified stars}
In the present sample, 58 variables are found to be field and unclassified population lying towards the direction of Stock 8. These variables have periods ranging from 0.051 d to 0.549 d.
The period and amplitude of star V31 which belongs to unclassified stars are 0.113 d and 0.042 mag, respectively. The light curve of V31 is
found to be similar to that of $\delta$ Scuti type variables. 
The probable BP variables namely stars V5, V16, V21, V24, V39, V40, V41, V44, V47, V52, V53, V54, V62, V63, V65, V67, V71, V72, V108, V112, and V124 
have period and amplitude in the range of 0.093 d to 0.378 d and 0.011 to 0.123 mag, respectively. The characteristics
of these stars as well as their light curves indicate that they could be $\beta$ Cep or SPB stars.

\begin{figure}
\includegraphics[width=8cm, height=8cm]{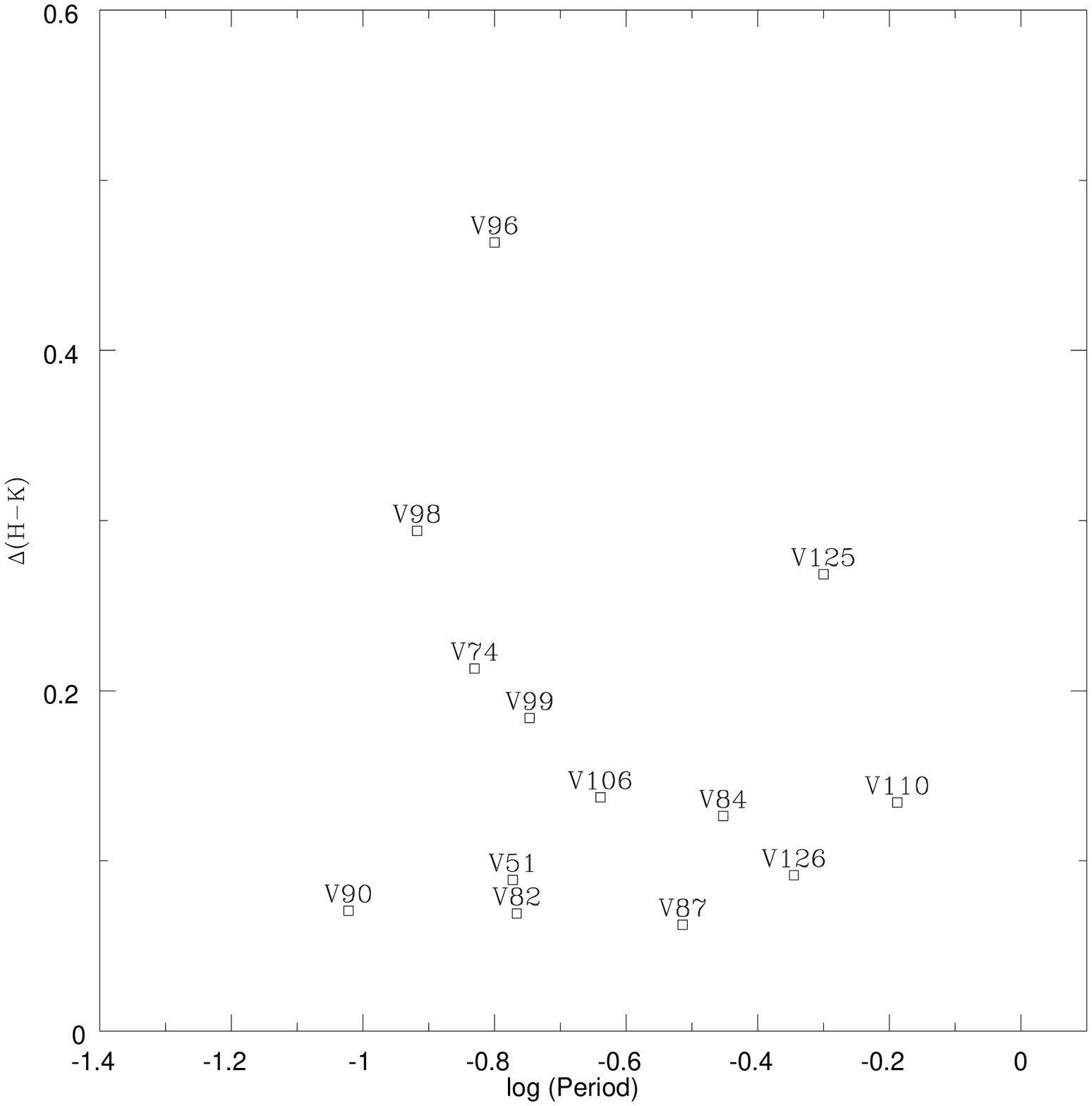}
\caption{$\Delta$$(H-K)$ vs rotation period.}
\end{figure}

\section{Rotation and $\Delta (H-K)$ }
In the case of PMS stars, it is suggested that the magnetic star-disk interaction could drive the angular momentum of the star (K\"onigl 1991).
The loss of angular momentum due to accretion-related processes suggests that accreting stars should be slower rotators than non-accreting ones.
A few studies have hinted to a correlation between the rotation rate and the NIR/IR excess of PMS objects (Edwards et al. 1993, Lamm et 
al. 2005, Rebull et al. 2006, 2014, Lata et al. 2016). However, Stassun et al. (1999) in the case of Orion Nebula Cluster (ONC) and Cieza \& Baliber (2006) in the
case of IC 348 did not find any correlation between accretion and rotation.

The conflicting results, whether a correlation exists between rotation and accretion process in PMS or not, may be due to several causes e.g. mass of PMS stars; statistical
robustness of the data etc. could be some of the reasons. 
Cieza \& Baliber (2007) have shown that if effect of mass is removed, a clear
trend in  the disk fraction with period can be observed in NGC 2264 and Orion 
nebula clusters. Venuti et al. (2015, 2016) have found that the connection between 
rotational properties and accretion traced via UV excess measurements is
consistent with earlier findings (based on IR excess measurements) in the 
sense that fast rotators are typically devoid of dusty disks. 

The data in the present work are very small, however, a plot between NIR excess index, $\Delta (H-K)$ and period in 
Fig. 14 indicates no correlation between $\Delta (H-K)$  and period for short period variables. 
 Where $\Delta (H-K)$ is the horizontal displacement of the YSO location from the left reddening vector
of `F' region in the $(J-H)/(H-K)$ TCD (Fig. 5).
It is worthwhile to mention here that the plot between UV excess and rotation
period by Venuti et al. (2016, see figure 11) does not show any
correlation between UV excess and period for PMS stars having period $\lesssim$ 1 day, which is consistent with the present results.

\section{Summary}
In the current study, we have presented time series photometry of 130 variables in the field of young open cluster Stock 8. We identified 51, 50 and 29 stars as members of the cluster, non members and unclassified, respectively. Member stars are categorized as 28 MS and 23
PMS stars. Twenty one of 50 variables identified as field star population could belong to BP population. 
Of 28 MS variables, 1, 2, 7 and 14 stars are classified as $\beta$ Cep, $\delta$  Scuti, SPB, and new class variables, respectively.
Since the variability behaviour for 14 new class variables is not
easily explained, additional photometric monitoring of these objects
is desirable. Four MS variables remain uncharacterized in the present work.
Five and 2 of 23 PMS variables are found to be
CTTSs and HAeBe stars, respectively, whereas 16 PMS stars are found to be WTTSs.
 For the majority of these
PMS variables, the ages are found to be $\lesssim$ 5 Myr while masses are in the range from 0.50 to 2.8 $M_{\odot}$.
Based on its light curve shape, period and amplitude, 
star V31 could be classified as a $\delta$ Scuti variable.
No correlation between $\Delta (H-K)$  and rotation period is found for short period PMS variables.

\section{Acknowledgment}
Authors are thankful to the anonymous referee for useful scientific suggestions/comments.
This work has made use of data obtained at the Thai National Observatory
on Doi Inthanon operated by NARIT.
AKP is thankful to the NARIT for providing support during his visit to NARIT. 

\section*{References}
\noindent
Balona L. A., Pigulski A., Cat P. De, Handler G., Guti$\acute{e}$rrez-Soto J., Engelbrecht C. A., Frescura F., Briquet M., et al., 2011, MNRAS, 413, 2403 \\
Bessell M. S., Brett J. M., 1988, PASP, 100, 1134 \\
Bouvier J.  et al., 1997, A\&A, 318, 495 \\
Carpenter J. M., Hillenbrand L. A., Skrutskie M. F., 2001, AJ, 121, 3160 \\
Chauhan N., Pandey A. K., Ogura K., Ojha D. K., Bhatt B. C., Ghosh S. K., Rawat P. S., 2009, MNRAS, 396, 964 \\
Cieza L., Baliber N.,  2006, ApJ, 649, 862 \\
Cieza L., Baliber N., 2007, ApJ, 671, 605 \\
Cohen J. G., Persson S. E., Elias J. H., Frogel J. A., 1981, ApJ, 249, 481 \\
Cutri R. M., et al . VizieR Online Data Catalog: WISE All-Sky Data Release, VizieR On-line Data Catalog: II/311. Vol. 2311. 2012. p. 0. \\
Cutri R. M., Skrutskie M. F., van Dyk S., Beichman C. A., Carpenter J. M., Chester T., Cambresy L., Evans T., Fowler J., Gizis J., et al., 2003, 2MASS All Sky Catalog of point sources \\
Dhillon V. S., Marsh T. R., Atkinson D. C., Bezawada N., Bours M. C. P., Copperwheat C. M., Gamble T., Hardy L. K., Hickman R. D. H., Irawati P., Ives D. J., Kerry P., Leckngam A., Littlefair S. P., McLay S. A., O'Brien K., Peacocke P. T., Poshyachinda S., Richichi A., Soonthornthum B., Vick A., 2014, MNRAS, 444, 4009 \\
Duquennoy A. and Mayor M., 1991, A\&A, 248, 485 \\
Edwards S., Strom S. E., Hartigan P., Strom K. M., Hillenbrand L. A., Herbst W., Attridge J., Merrill K. M., Probst R., Gatley I., 1993, AJ, 106, 372 \\
Girardi L., Bertelli G., Bressan A., Chiosi C., Groenewegen M. A. T., Marigo P., Salasnich B., \& Weiss A., 2002, A\&A, 391, 195 \\
Grankin K. N., Melnikov S. Yu., Bouvier J., Herbst W., Shevchenko V. S., 2007, A\&A, 461, 183 \\
Grankin K. N., Bouvier J., Herbst W., Melnikov S. Y, 2008, A\&A, 479, 827 \\
Guti$\acute{e}$rrez-Moreno A.,  PASP, 1975, 87 805 \\
Herbst W., Herbst D. K., Grossman E. J., Weinstein D., 1994, AJ, 108, 1906 \\
Jose J., Pandey A. K., Ojha D. K., Ogura K., Chen W. P., Bhatt B. C., Ghosh S. K., Mito H., Maheswar G., Sharma S., 2008, MNRAS, 384, 1675 \\	
Jose J. Herczeg G. J., Samal M. R., Fang Q., Panwar, N., 2017, ApJ, 836, 98 \\
Joy A. H., 1945, ApJ, 102, 168 \\
K\"onigl A., 1991, ApJ, 370, 39 \\
Lamm M. H., Mundt R., Bailer-Jones C. A. L., Herbst W., 2005, A\&A, 430, 1005 \\
Lata S., Pandey A. K., Maheswar G., Mondal S., Kumar B., 2011, MNRAS, 418, 1346 \\
Lata S., Pandey A. K., Chen W. P., Maheswar G., Chauhan N., 2012, MNRAS, 427, 1449 \\
Lata S., Yadav Ram Kesh, Pandey A. K., Richichi A., Eswaraiah C., Kumar B., Kappelmann N., Sharma S., 2014, MNRAS, 442, 273 \\
Lata S., Pandey A. K., Panwar Neelam, Chen W. P., Samal M. R., Pandey J. C., 2016, MNRAS, 456, 2505 \\
Lomb N. R., 1976, ApSS, 39, 447 \\
Menard F., Bertout C, 1999, in Lada C. J., Kylafis N. D., eds, The Origin of Stars and Planetary Systems. Kluwer Academic Publishers, Dordrecht, p. 341 \\
Meyer M. R., Calvet N., Hillenbrand L. A., 1997, AJ, 114, 288 \\
Mowlavi N., Barblan F., Saesen S., Eyer L., 2013, A\&A, 554, 108 \\
Pandey A. K., Sharma S., Ogura K.,  2006, MNRAS, 373, 255 \\
Pandey A. K., Samal M. R., Yadav R. K., Richichi A., Lata S., Pandey J. C., Ojha, D. K., Chen W. P.,  2014, NewA, 29, 18 \\
Percy J. R., Esteves S., Glasheen J., Lin A., Long J., Mashintsova M., Terziev E., Wu S., 2010, J. Am. Assoc. Var. Star Observers, 38, 151 \\
Richichi A., Irawati P., Soonthornthum B., Dhillon V. S., Marsh T. R., 2014, AJ, 148, 100 \\
Rebull L. M., Stauffer J., Megeath S., Hora J. L., Hartmann L., 2006, ApJ, 646, 297 \\
Rebull L. M., Cody A. M., Covey K. R., et al., 2014, AJ, 148, 92 \\
Robitaille T. P., Whitney B. A., Indebetouw R.,  Wood K.,  2007, ApJS, 169, 328 \\
Robitaille T. P., Whitney B. A., Indebetouw R., Wood K.,  Denzmore P.,  2006, ApJS, 167, 256 \\
Samal M. R., Pandey A. K., Ojha D. K., et al., 2010, ApJ, 714, 1015 \\
Scargle J. D., 1982, ApJ, 263, 835 \\
Sesar B. et al, 2007, AJ, 134, 2236 \\
Siess L., Dufour E., Forestini M., 2000, A\&A, 358, 593 \\
Stetson P.~B., 1987, PASP, 99, 191
Stetson P.~B, 1992, J. R. Astron. Soc. Can., 86, 71 \\
Stassun Keivan G., Mathieu R. D., Mazeh T., Vrba F. J., 1999, AJ, 117, 2941 \\
Torres G.,  AJ, 2010, 140, 1158 \\
Venuti L,  Bouvier J., Irwin J., Stauffer J. R., Hillenbrand L. A., Rebull L. M., Cody A. M., Alencar S. H. P., Micela G., Flaccomio E., Peres G., 2015, A\&A, 581, 66 \\
Venuti L., Bouvier J., Cody A. M., Stauffer J. R., Micela G., Rebull L. M., Alencar S. H. P., Sousa A. P., Hillenbrand L. A., Flaccomio E., 2016, arXiv161008811V \\
Waelkens C., 1991, A\&A, 246, 453 \\

\end{document}